\input epsf
\font\fourteenbf=cmbx12 at 14pt
\font\fourteenrm=cmr12 at 14pt
\font\fourteenit=cmti12 at 14pt
\font\twelvebf=cmbx12
\font\twelverm=cmr12
\font\twelveit=cmti12
%\font\twelveex=cmex12
\font\twelvesl=cmsl12
\font\tenbf=cmbx10
\font\tenrm=cmr10
\font\tenit=cmti10
\font\tenex=cmex10
\font\tensl=cmsl10
\font\ninebf=cmbx9
\font\ninerm=cmr9
\font\nineit=cmti9
\font\nineex=cmex9
\font\ninesl=cmsl9
\font\eightbf=cmbx8
\font\eightrm=cmr8
\font\eightit=cmti8
\font\eightex=cmex8
\font\eightsl=cmsl8
\font\sevenbf=cmbx7
\font\sevenrm=cmr7
\font\sevenit=cmti7
\font\sevenex=cmex7
%\font\sevensl=cmsl7

\def\fourteen{\let\rm=\fourteenrm \let\bf=\fourteenbf \let\it=\fourteenit
            \let\sl=\fourteensl}
\def\twelve{\let\rm=\twelverm \let\bf=\twelvebf \let\it=\twelveit 
            \let\sl=\twelvesl}
\def\ten{\let\rm=\tenrm \let\bf=\tenbf \let\it=\tenit
         \let\ex=\tenex \let\sl=\tensl}
\def\nine{\let\rm=\ninerm \let\bf=\ninebf \let\it=\nineit
          \let\ex=\nineex \let\sl=\ninesl}
\def\eight{\let\rm=\eightrm \let\bf=\eightbf \let\it=\eightit
           \let\ex=\eightex \let\sl=\eightsl}
\def\seven{\let\rm=\sevenrm \let\bf=\sevenbf \let\it=\sevenit
           \let\ex=\sevenex } 

%\TagsOnRight
\hsize=6.0truein
\vsize=8.6truein

%\nopagenumbers
\baselineskip=15pt
\parindent=15pt
\twelve

%%%%%%%%%%%%%%%%%%%%%%%%%%%%%%%%%%%%%%%%%%%%%%%%%%%%%%%%%%%%%%%%%%%%%%
%Selectable parameters
\newif\ifdebug % True => Print Eq name
\newif\ifbacketedref % True => ref are [n] else n
\newif\ifpareqref % True => Equation ref are (n) else Eq. n
\newif\ifpareqno % True => In Equation, No (n) else n
\newif\ifeqnoperchapter % True => Equation numbered per chapter : C.e
\def\chapterfont{\twelvebf}

\def\titlefont{\twelvebf}
%%%%%%%%%%%%%%%%%%%%%%%%%%%%%%%%%%%%%%%%%%%%%%%%%%%%%%%%%%%%%%%%%%%%%%
\def\centerline#1{\line{\hss{#1}\hss}}%
\def\title#1{{\titlefont\centerline{#1}\vskip 5mm}}
\def\authors#1{\centerline{#1} }
\def\DURHAM{{\parskip 2mm\tenit\baselineskip=13pt
    \centerline{Department of Mathematical Sciences}
    \centerline{University of Durham, Durham DH1 3LE, England}
}}

\def\Email#1{{\parskip 2mm\tenit\baselineskip=13pt\centerline{E-Mail: #1}}}

\def\abstract#1{{\tenrm\baselineskip=13pt
\parindent=0pt\vskip 1cm
\centerline{ABSTRACT}
\vbox{}
\vbox{\leftskip=12mm\rightskip=12mm#1}
}}
\newtoks\date 
\def\monthname{\relax\ifcase\month 0/\or January\or February\or
    March\or April\or May\or June\or July\or August\or September\or
    October\or November\or December\else\number\month/\fi}
\date={\monthname\ \number\day, \number\year}
\def\pubnum#1{{\twelverm\obeylines\everypar{\hfill} 
\rm DTP-#1
\the\date}
\vskip 5mm}
% CHAPTERS
%=========
\newcount\chapternumber \chapternumber=0 
\newcount\sectionnumber \sectionnumber=0
\def\chapterreset{%
   \global\advance\chapternumber by 1%
   \ifeqnoperchapter \global\equanumber=0 \fi
   \sectionnumber=0 
}
\def\chapter#1{
{ \parindent=0pt \vglue 0.5cm
  \baselineskip=13pt
  \chapterreset
  {\chapterfont \number\chapternumber .\ #1 \hfill}
  \vglue 0.2cm
}}

%%%%%%%%%%%%%%% EQUATIONS %%%%%%%%%%%%%%%%%%%%%
\newcount\equanumber         \equanumber=0
\def\eqname#1{
  \def\Ename{\string#1}
  \relax 
  \ifnum\the\equanumber<0
    \xdef\LEQno{{\tenrm(\number-\equanumber)}}\global\advance\equanumber by -1
  \else \global\advance\equanumber by 1
    \ifdebug
       \xdef\LEQno{\Ename} 
    \else
      \ifeqnoperchapter
        \xdef\LEQno{\number\chapternumber .\number\equanumber} 
      \else
        \xdef\LEQno{\number\equanumber} 
      \fi
    \fi
  \fi
  \ifpareqref
    \xdef#1{{\rm(\LEQno)}\ }
  \else
    \xdef#1{{\rm Eq.\LEQno}\ }
 \fi
%write equation No (par or not)  
  \ifpareqno {\tenrm(\LEQno)} \else {\tenrm\LEQno} \fi 
}

\def\eqn#1{\eqno\eqname{#1}}

%%%%%%%%%%%%%%%%%% REFERENCES %%%%%%%%%%%%%%%%%%%%%%%
\newif\ifbacketedref
\newif\ifreferenceopen       \newwrite\referencewrite
\newdimen\referenceminspace  \referenceminspace=25pc
\newdimen\refindent          \refindent=30pt
\newcount\referencecount     \referencecount=0
\def\reffile{rf}
\immediate\openout\referencewrite=\reffile
\def\refout{\par \penalty-400 \vskip\chapterskip %\spacecheck\referenceminspace
 \immediate\closeout\referencewrite
   \referenceopenfalse
   \vglue 0.5cm
   \baselineskip=13pt
   {\chapterfont References\hfil}
   \vglue 0.2cm
   \baselineskip=15pt
   \input \reffile
   }
\def\Textindent#1{\noindent\llap{#1\enspace}\ignorespaces}
\def\refitem#1#2{\par \hangafter=0 \hangindent=\refindent \Textindent{#1} #2}
\def\refnum#1{\global\advance\referencecount by 1 \def\Rname{\string#1}
\ifdebug\xdef#1{\Rname}\else\xdef#1{\the\referencecount}\fi
}
\def\refmark#1{\hbox{\raise1ex\hbox{{\eightrm%
   \ifbacketedref[#1]\else#1\fi}}}}

\def\Ref#1#2{\refnum#1\refmark{#1}%
  \immediate\write\referencewrite{\noexpand\refitem{#1.}{#2}}%
}
\def\REF#1#2{\refnum#1%
 \immediate\write\referencewrite{%
 \noexpand\refitem{#1.}{#2}}%
}

\newskip\headskip 	\headskip=8pt plus 3pt minus 3pt
\newskip\chapterskip    \chapterskip=\bigskipamount
\newskip\sectionskip     \sectionskip=\medskipamount
\def\ack{\par\penalty-100\medskip 
    \line{\twelverm\hfil ACKNOWLEDGEMENTS\hfil}\nobreak\vskip\headskip }

%%%%%%%%%%%%%%%%%%%%%%% TABLES %%%%%%%%%%%%%%%%%%%%%%%%%%
\newcount\Tableno     \Tableno=0
\def\figureinc{%
   \global\advance\figureno by 1%
}
%% Table{Name}{Caption}{The Table}
\def\Table#1#2#3{\vskip 10mm%
{#3}
\global\advance\Tableno by 1%
\xdef#1{{\rm\number\Tableno}\ }%
\vskip 5mm
\advance\rightskip by 1cm
\vbox {\par\parindent=1cm\hangindent=28mm\hangafter=1
Table \number\Tableno\ : #2 }
\advance\rightskip by -1cm
\vskip 10mm
}
%% NextTableNo : the table Numb for the next Table
\def\NextTableNo{{\advance\Tableno by 1 \number\Tableno}}
%
%%%%%%%%%%%%%%%%%%%%%%% FIGURES %%%%%%%%%%%%%%%%%%%%%%%%%%
% uses epsf.tex : add \input epsf at the top of the file
%%%%%%%%%%%%%%%%%%%%%%%%%%%%%%%%%%%%%%%%%%%%%%%%%%%%%%%%%%
\newcount\figureno     \figureno=0
\newdimen\figdim       \figdim=70mm
\def\figureinc{%
   \global\advance\figureno by 1%
}
\def\figcaption#1#2#3{\hbox to #2{\hss{\vbox{\hsize=#2 \parindent=0pt 
        {\bf Figure \number\figureno#3 :\ }#1}}\hss}
}

\def\OneFig#1#2{\vskip 5mm\figdim=70mm
\centerline{\figureinc
  \vbox{\epsfxsize=6cm\epsfysize=6cm\epsfbox{#1}\vskip 5mm
        \figcaption{#2}{\figdim}{}}
  }\vskip 5mm
}
\def\TwoFigs#1#2#3#4{\vskip 5mm\figdim=70mm
 \hbox to \hsize {
  \vbox {\figureinc\epsfxsize=7cm\epsfysize=7cm\epsfbox{#1}\vskip 5mm
         \figcaption{#2}{\figdim}{}
  }\hfill
  \vbox {\figureinc\epsfxsize=7cm\epsfysize=7cm\epsfbox{#3}\vskip 5mm
         \figcaption{#4}{\figdim}{}}}
 \vskip 5mm
}
\def\TwoFigsAB#1#2#3#4{\vskip 5mm\figdim=70mm
 \hbox to \hsize {\figureinc
  \vbox {\epsfxsize=7cm\epsfysize=7cm\epsfbox{#1}\vskip 5mm
         \figcaption{#2}{\figdim}{.a}
  }\hfill
  \vbox {\epsfxsize=7cm\epsfysize=7cm\epsfbox{#3}\vskip 5mm
         \figcaption{#4}{\figdim}{.b}}}
 \vskip 5mm
}
\def\FourFigsAD#1#2#3#4#5#6#7#8{\vskip 5mm\figdim=70mm\figureinc
 \hbox to \hsize {
  \vbox {\epsfxsize=7cm\epsfysize=7cm\epsfbox{#1}\vskip 5mm
         \figcaption{#2}{\figdim}{.a}
  }\hfill
  \vbox {\epsfxsize=7cm\epsfysize=7cm\epsfbox{#3}\vskip 5mm
         \figcaption{#4}{\figdim}{.b}}}
 \vskip 5mm
 \hbox to \hsize {
  \vbox {\epsfxsize=7cm\epsfysize=7cm\epsfbox{#5}\vskip 5mm
         \figcaption{#6}{\figdim}{.c}
  }\hfill
  \vbox {\epsfxsize=7cm\epsfysize=7cm\epsfbox{#7}\vskip 5mm
         \figcaption{#8}{\figdim}{.d}}}
 \vskip 5mm
}
%%%%%%%%%%%%%%%%%%% GENERAL %%%%%%%%%%%%%%%%%%%%

\def\mod#1{ \vert #1 \vert }

\def\Tanh{\hbox{tanh}}

\def\C {{\rlap{\kern 1.0mm \vrule height 7pt depth 0pt} \rm C}}
\def\R {{\rlap{\kern 0.1mm \vrule height 7pt depth 0pt} \rm R}}
\def\U {{\rlap{\kern 1.2mm \vrule height 7pt depth 0pt} \rm 1}}

% DEFAULT SETTING
%%%%%%%%%%%%%%%%%
\debugfalse
\backetedreftrue
\eqnoperchaptertrue
\pareqreftrue 
\pareqnotrue
\eqnoperchaptertrue

\twelve
\rm

%1
\def\rNo{B.M.A.G. Piette, B.J. Schroers and W.J. Zakrzewski, 
Z. Phys. C {\bf65} (1995) 165, Nucl. Phys. B {\bf 439} (1995) 205} 
%2
\def\rNt{B.M.A.G. Piette, H.J.W. Muller-Kirsten, D.H. Tchrakian and 
W.J. Zakrzewski, Phys.Lett. B {\bf 320} (1994) 294}
%3
\def\rNT{A. Kudryavtsev, B. Piette, W.J. Zakrzewski, 
Durham preprint DTP96/17, Z. Phys. C (1997) to be published.}
%4
\def\rNf{R.A. Leese, M. Peyrard and W.J. Zakrzewski,
%Soliton Scatterings in Some Relativistic Models  in (2+1) Dimensions 
Nonlinearity {\bf 3} (1990) 773}
%5
\def\rNF{ I.L. Bogolyubski, A.A. Bogolyubskaya, Phys Lett. 
{\bf 395} (1977) 269}
%6
\def\rNs{ Yu-dong Yin, Tao Huang, Jiang-ru Wen. Phys Lett. 
{\bf 373} (1996) 309}
%7
\def\rNS{ M.B. Voloshin. Sov. J. Nucl. Phys. 
{\bf 21} (1975) 687}
%8
\def\rNe{ G.R. Farrar, J.W. McIntosh, Jr. Phys Rev. 
{\bf D51} (1995) 5889}
%9
\def\rNn{ C.G. Callan Jr. and J.A. Harvey. Nuck Phys. 
{\bf B250} (1985) 427}
%10
\def\rNoz{ D.B. Kaplan, M. Schmaltz, Phys Lett.  
{\bf B368} (1996) 44}
%11
\def\rNoo{ A.G. Cohen, D.B. Kaplan and A.E. Nelson, 
Ann Rev of Nucl and Part Sci, {\bf 43} (1993) 27}
%12
\def\rNot{ P. Huet, A.E. Nelson, Phys. Lett. 
{\bf B355} (1995) 229}
%13
\def\rNoT{ Vachaspati and T. Vachaspati. 
Phys. Lett.{\bf B238} (1990) 41}
%14
\def\rNof{ J. Dziarmaga Phys Lett  
{\bf B328} (1994) 392}
%15
\def\rNoF{ A.M. Kosevich, B.A. Ivanov and A.S. Kovalev. 
Phys Reports, {\bf 194} (1990) 1}
%16
\def\rNos{ J. Pouget and G.A. Maugin. Phys Rev.
{\bf B30} (1984) 5306}

%10

\pubnum{97/25}

\title{Skyrmions and domain walls in (2+1) dimensions}
\authors{A. Kudryavtsev$\sp{1}$}
\authors{B.M.A.G Piette,}
\authors{and}
\authors{W.J. Zakrzewski}
\DURHAM
\centerline{$\sp{1}$ also at ITEP, Moscow, Russia}
\Email{Kudryavtsev@vitep5.itep.ru \quad B.M.A.G.Piette@uk.ac.durham\quad W.J.Zakrzewski@uk.ac.durham}
 
\abstract{We study classical solutions of the vector $O(3)$ sigma model
in $(2+1)$ dimensions, spontaneously broken to $O(2) \times Z_2$.
The model possesses Skyrmion-type solutions as well as stable domain walls 
which connect different vacua. We show that different types of 
waves can propagate on the wall, including waves carrying a topological charge.
The domain wall can also absorb Skyrmions and, under appropriate
initial conditions, it is possible to emit a Skyrmion from the wall.
}

\chapter{Introduction.}
\REF\RNo{\rNo} \REF\RNt{\rNt}  \REF\RNT{\rNT}
In our previous works [\RNo-\RNT]
various solutions of the so-called baby-Skyrmion model were investigated. In 
this paper we begin
a  systematic study of solutions of the nonlinear vector $O(3)$ sigma model
spontaneously broken to $ O(2) \times Z_2$.

The $(2+1)$-dimensional version of this model is described by 
the Lagrangian density
 
$$ 
L=F_{\pi}\bigl({{1\over2}\partial_{\alpha}\vec \phi \partial^{\alpha} 
\vec \phi - 
{k^{2}\over4}
(\partial_ {\alpha} \vec \phi \times \partial_{\beta}\vec \phi )
(\partial^{\alpha} \vec \phi \times \partial^{\beta} \vec \phi)-
 {\mu^{2}\over 2}   (1-\phi^2_3)\bigr)}.
\eqn\eLagPhi
$$      

Here $\vec \phi \equiv (\phi_{1},\phi_{2},\phi_{3}) $ denotes a triplet 
of scalar real 
fields which satisfy the constraint ${\vec \phi}^{2}=1$;
$( \partial_{\alpha} \partial^{\alpha}=\partial_{t}\partial_{t}-\partial_{x} 
\partial_{x}-\partial_{y}\partial_{y})$.

\REF\RNf{\rNf} 
The first term in \eLagPhi is the familiar Lagrangian density of the pure
$S\sp2$ $\sigma$-model. The second term, fourth order in derivatives, is the 
(2+1) dimensional analogue of the Skyrme-term of the three-dimensional 
Skyrme-model. The last term is often referred to  as a potential term.  
The last two terms in the Lagrangian \eLagPhi
are added to guarantee the stability of a Skyrmion [\RNf]. In 
contradistinction to 
the first two terms the potential term in \eLagPhi differs from that of the 
baby-Skyrmion model [\RNo-\RNT] and is quadratic in terms of the field 
$\phi_3$. 
Our potential looks like a mass term for the $\phi_1$ and $\phi_2$ fields
but, unlike the baby-Skyrmion potential, it does not provide any additional 
interaction between them.  
Its normalisation was chosen so that in the limit of $\phi_3$
being close to $1$ both the baby-Skyrmion potential, $\mu^{2} (1-\phi_3)$, 
and our potential are identical and approximated by 
${\mu^{2} \over 2}(\phi_1^2 + \phi_2^2)$. 

As in [\RNo-\RNT] we fix our units of energy and length by setting  
$F_{\pi}=k=1$ and we choose $\mu^{2}=0.1$ to compare our results with what
was observed in [\RNo].

\REF\RNF{\rNF} \REF\RNs{\rNs} 
Notice, that the gauge version of the Lagrangian \eLagPhi was recently 
studied in [\RNF] (without the Skyrme term). Some aspects of the soliton 
stability for the model \eLagPhi were also discussed in [\RNs].

\REF\RNoF{\rNoF} \REF\RNos{\rNos}
Let us add that our model ( corresponding to the Lagrangian 
density  \eLagPhi ) can be 
derived in the continuous limit of easy-axis Heisenberg antiferromagnetics 
[\RNoF] and ferroelectrics with an easy-axis anisotropy [\RNos] (without 
the Skyrme term).

One clearly sees from \eLagPhi\ that the model has two distinct vacua:
$\phi_3=\phi_{\pm}=\pm 1$. If we require that the field at spatial infinity 
describes one of the two vacua, we can compactify the two dimensional plane 
into a sphere making the theory topological. All classical 
configurations will be characterised by the value of the winding number  
$$
N= {1 \over 8\pi } \int \varepsilon_{ij} \vec \phi
\bigl (\partial^i \vec \phi \times \partial^j \vec \phi \bigr ) d^2x
\eqn\ewindnum
$$
which is an integer as it counts the number of times the target sphere 
$S^2_{iso}$ is covered by the physical space $S^2_{space}$.

As we will demonstrate our model possesses two types of solutions, i.e. 
Skyrmions, and domain walls represented by a (1+1) dimensional soliton. Domain 
walls connect different vacua $\phi_{\pm}$ of the theory making the 
interaction between a Skyrmion and the domain walls nontrivial.

\REF\RNS{\rNS} \REF\RNe{\rNe} \REF\RNn{\rNn} \REF\RNoz{\rNoz}
\REF\RNoo{\rNoo} \REF\RNot{\rNot}
The interaction of particles with domain walls has become the subject of 
interest beginning with the paper of Voloshin [\RNS]. A detailed analysis 
of the  scattering of gauge Abelian particles on domain walls may be found in 
a recent paper by Farrar et al. [\RNe]. The physics of fermion zero modes on 
strings and domain walls was discussed widely in literature, especially 
in connection with the problem of anomalies, see e.g. [\RNn] and refs. 
therein. The problem of the chiral fermion determinant in the presence of a 
domain wall was discussed recently in [\RNoz]. It is also worth mentioning 
that the expanding bubbles of a new phase during the electroweak phase 
transition may be  an additional source for the CP-violation effects and 
for the electroweak baryogenesis, see e.g. [\RNoo,\RNot].

In this paper we consider a slightly different aspect of the interaction of 
particles with domain walls. Namely, we study the interaction of two objects 
of our model, i.e. a Skyrmion and a domain wall. We demonstrate that this 
interaction is nontrivial and that it leads to the absorption of the Skyrmion 
by the wall. The conserved topological charge of the Skyrmion is transferred 
to the waves on the wall which then carry a topological charge. Under 
certain initial conditions the collision of the waves of the wall stimulates 
the emission of an isolated Skyrmion off the wall.

\chapter{Travelling waves on the domain wall}

First of all, we will study soliton like solutions of the theory 
\eLagPhi. To do this it is convenient to change variables and to use 
two real fields $f$ and $\psi$ instead of $\vec \phi$,

$$
\vec \phi=\bigl (\sin f \cos \psi, \sin f \sin \psi, \cos f \bigr)
\eqn \ePhiFPsi
,$$
where $f \in [0,\pi]$ and $\psi \in [0,2\pi]$. In terms of $f$ and $\psi$ the 
Lagrangian  \eLagPhi has the following form

$$
L=
{1\over 2} \bigl (f_t^2-f_x^2-f_y^2 \bigr ) +
{\sin^{2}f \over 2} \bigl [\bigl (f_t \psi_x-f_x \psi_t \bigr )^2 + \bigl 
(f_t \psi_y-f_y \psi_t \bigr )^2-\bigl (f_x \psi_y-f_y \psi_x \bigr )^2 \bigr ]
-{\mu^{2} \over 2} \sin^2f.
\eqn \eLagFPsi
$$

First we look for one dimensional static solutions of the equations of motion 
for \eLagFPsi in the form $f=f(x),\, \psi=const$. Then the equation for $f$ 
reduces to
$$
f_{xx}-{\mu^2 \over 2} \sin{2f}=0,
\eqn \eFxxsin
$$
which is a static version of the sine-Gordon equation with the  
soliton solution
$$
f_s(x)=2\arctan \exp \bigl ( \pm \mu (x-x_0)\bigr ).
\eqn \eSinSol
$$
This solution links the two vacua states $\phi_{\pm}$ of the theory and looks 
like a wall in the $(x,y)$-plane, which separates domains of different vacua. 
In contradistinction to the baby-Skyrmion model studied in previous 
papers [\RNo-\RNT], in the present model the soliton \eSinSol is stable, see 
also [\RNs]. Its energy is $E = 2 \mu = 0.2$ per unit length of the wall.

Notice that our soliton solution \eSinSol\ does not satisfy our boundary
conditions for the field at spatial infinity which we introduced to split all 
field configurations into integer valued classes labelled by the topological 
charge $N$. So it may appear that we cannot use the topological concepts
when looking at fields such as \eSinSol. However, this is not the case. We can
always consider the solution \eSinSol\ as having come, cut out and expanded, 
from a field configuration in the shape of a long straight wall whose ends 
are connected by a semicircular ring of an appropriate large radius. 

Using the expression \ewindnum\ and restricting the integration to a region of 
space we can define the topological charge for the field in a given region.
Then, calculating the topological charge for the straight part of the wall we 
find that it is zero. We will show later that the wall \eSinSol\ can be
modified so as to carry a non-zero topological charge.

%The boundary condition for \eSinSol\ is not compatible with the boundary
%condition we had to introduce to make to model topological as the value of
%the field at infinity depends on the region of space. Nevertheless we can 
%motivate this solution by thinking of the two ends of the wall being
%connected together to form a very large ring. We would expect such a ring to 
%shrink or expand slowly but the curvature of the ring is very 
%small and locally it can be well approximated by a straight wall. Notice also 
%that the overall topological charge of such a ring is one.

The domain wall acts like a wave carrier. To study waves carried by the wall
we start by looking at infinitesimal perturbation of the solution  
\eSinSol of the form 
$$
f(x,y,t)=f_s(x)+g(x,y,t)
\eqn\eSmallKink
$$ 
where $\mod {g(x,y,t)} \ll 1$ and where $f$ is 
given by \eSinSol. The linearised equation of motion for $g(x,y,t)$ is
$$
g_{tt}-g_{xx}-g_{yy}+\mu^2 \bigl (1-{2\over cosh^2(\mu x)} \bigr )g=0.
\eqn \eGxxNul
$$
We can look for solutions of the form 
$g(x,y,t)=\cos(\omega t-ky)\rho(x)$, which correspond to waves propagating
along the wall. When $\omega^2=k^2$ it is more convenient to write the
solutions in the form
\REF\RNoT{\rNoT}\REF\RNof{\rNof}
$$
g_0(x,y,t)=\mu \bigl[ F_1(y-t)+F_2(y+t)\bigr]\cosh^{-1}(\mu x),
\eqn \eGzero
$$ 
where $F_1(z)$ and $F_2(z)$ are arbitrary dimensionless functions of their 
arguments ($\mod{g_0} \ll 1$).
We call such fields zero modes because the $x$ dependence of such solutions
corresponds to the zero mode of the sine-Gordon kink. Putting together
\eSmallKink\ and \eGzero\ we can write 
$$
f(x,y,t)=f_s(x)+ \bigl[ F_1(y-t)+F_2(y+t)\bigr] 
{\partial f_s \over \partial x}
\eqn\eSmallKinkTwo
$$
showing that the additional waves correspond to a small displacement of the 
wall along the transversal ($x$) direction and which propagate at the speed of 
light along the wall. 
These solutions, called travelling waves, were found in [\RNoT]. A more 
general form of the travelling waves in the presence of multivortex 
configurations was discussed in [\RNof]. 

Nonzero modes for the equation \eGxxNul can also be found. They exist for 
$\omega^2 \geq k^2+\mu^2$ and correspond to the excitations of the continuum 
around the soliton \eSinSol . In terms of waves in the $y$-direction these 
excitations look like massive particles. Notice that our model possesses an 
energy gap between light-like zero modes  and massive excitations.

Zero mode excitations of the wall can also be found outside the small 
amplitude 
limit $\mod g  \ll 1.$ To see this, we rewrite the equation of motion in 
terms of the function $f(x,y,t)\,\,(\psi=const)$:
$$
f_{tt}-f_{xx}-f_{yy}+{\mu^2 \over 2} \sin2f =0.
\eqn \eFttSin
$$
The zero-mode solutions of this equation are of the form:
$$
f^{(0)}_{\pm}(x,y,t)=f_s(x-x_0^{\pm}(y\pm t)) ,
\eqn \eFzero
$$
where $x_0^{\pm}(z)$ are arbitrary functions of their arguments.
The wave \eFzero\ represents a transversal displacement of the wall which 
travels with the velocity of light along the $y$ direction.
However, unlike in the 
linearised case (solutions \eGzero\ of equation \eGxxNul), the superposition 
of solutions $f^{(0)}_+$ and $f^{(0)}_-$ is not a solution of \eFttSin .

The interaction between two deformation waves is thus non-linear. We have
solved numerically the full equation of motion derived from \eLagPhi\
to analyse the scattering between 2 kink-like deformations travelling on 
the wall in opposite directions. The result is shown on Figure 1.
One can see from the figures that the collision process is followed by 
the assymetric emission of waves in the $(x,y)$ plane

%%%%%%%%%% FIGURE 1 %%%%%%%%%%%%%%
\FourFigsAD{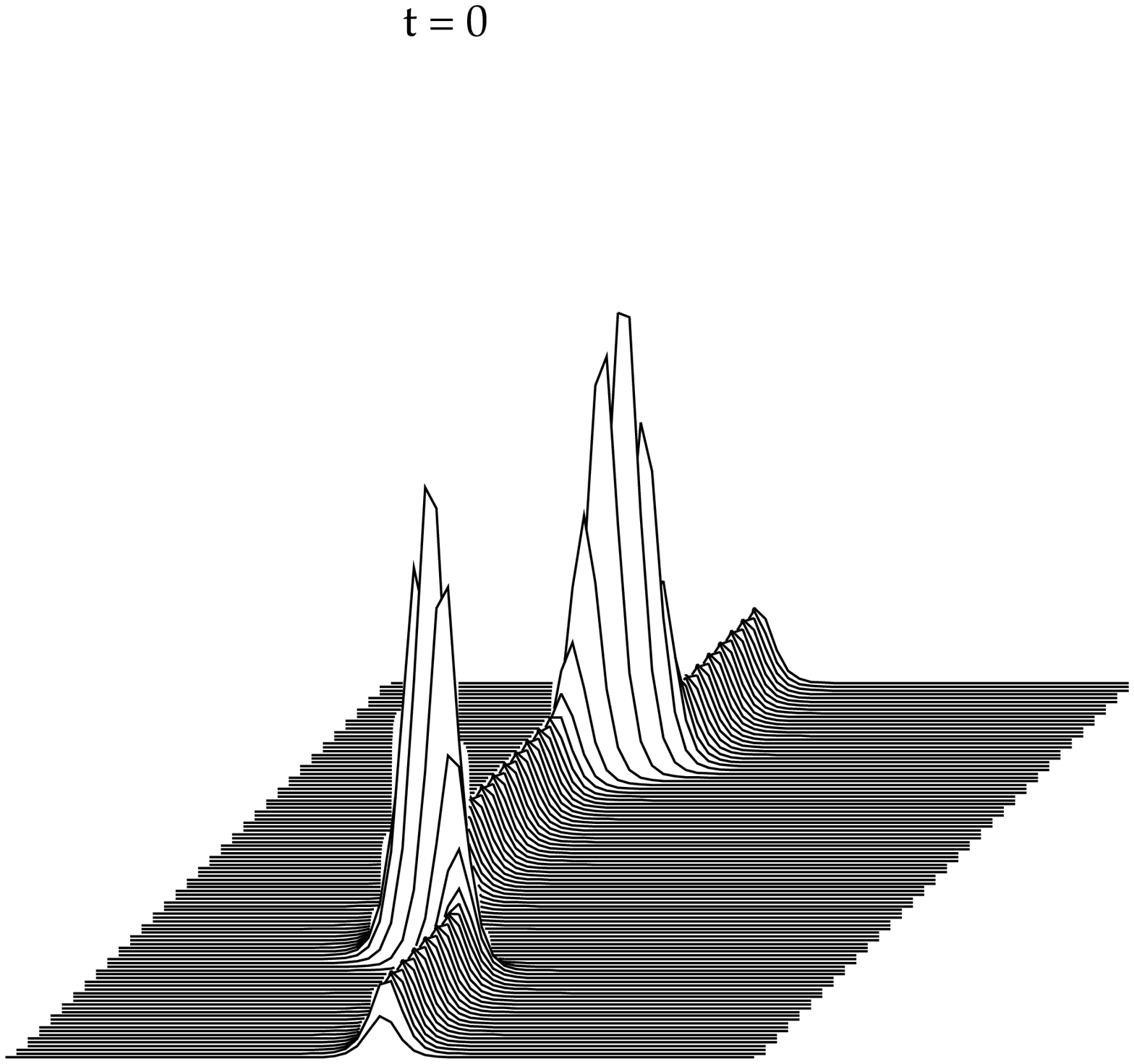}{Deformation wave : energy density at t = 0}{
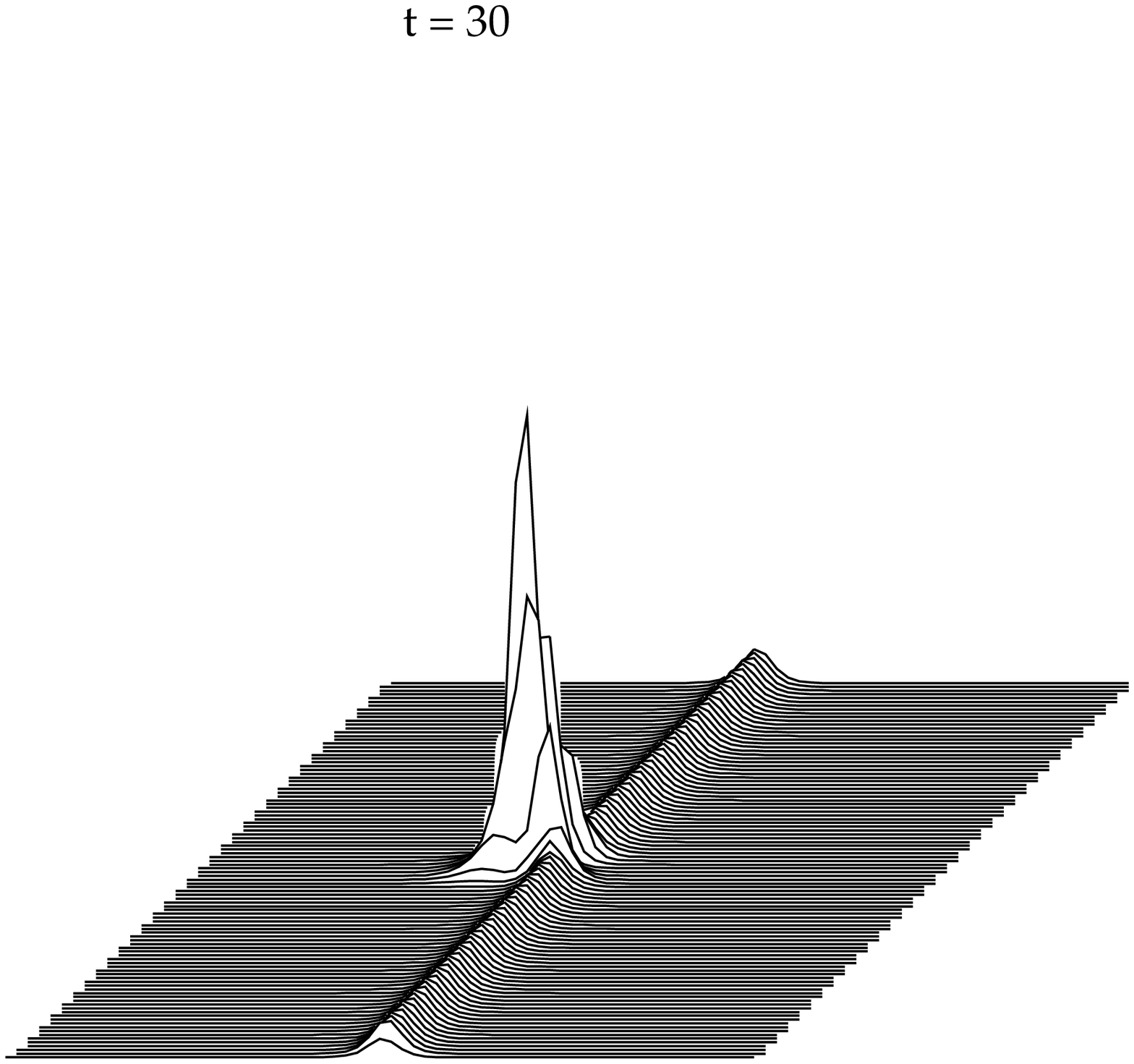}{Deformation wave : energy density at t = 30}{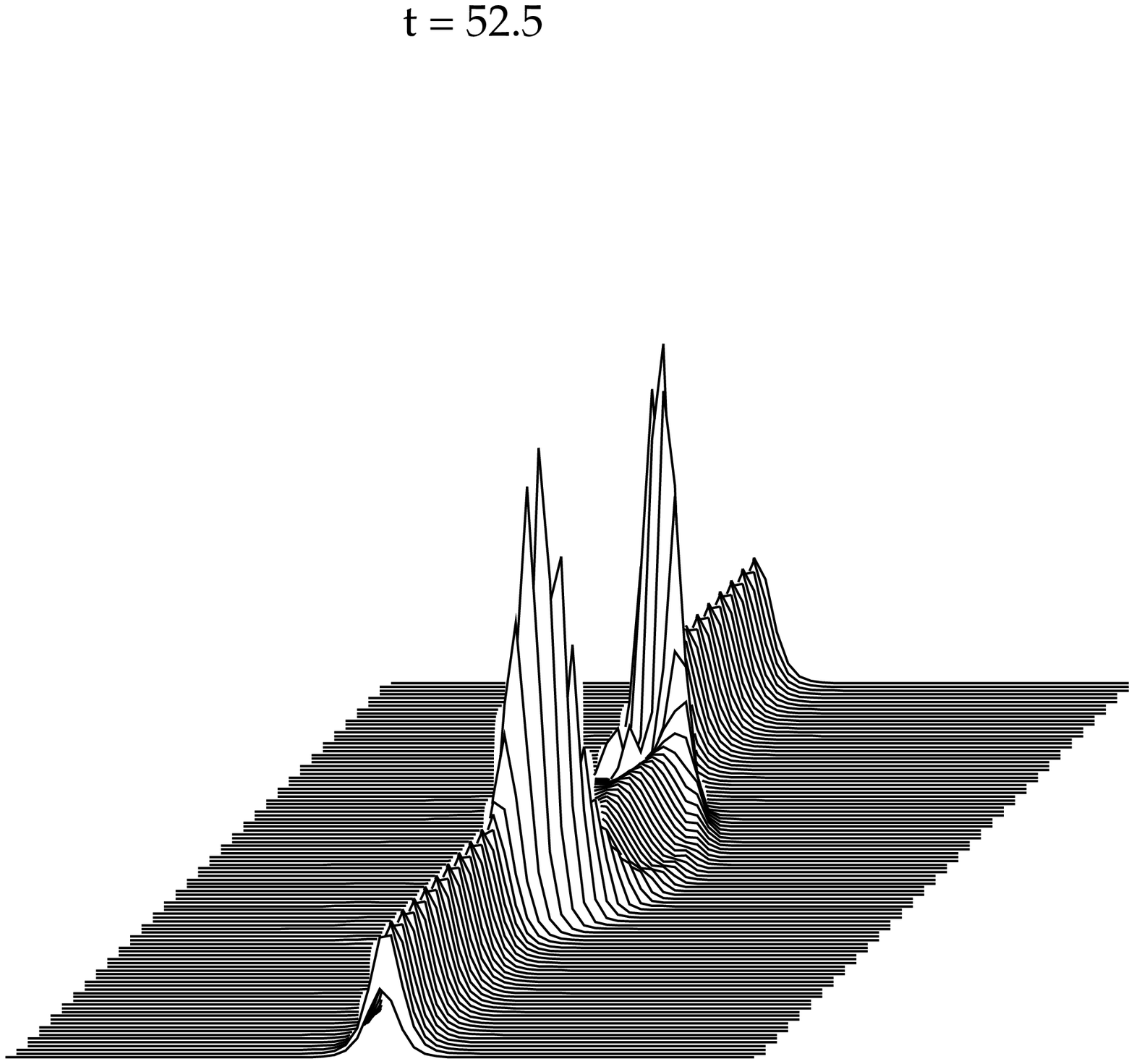}{
Deformation wave : energy density at t = 52.5}{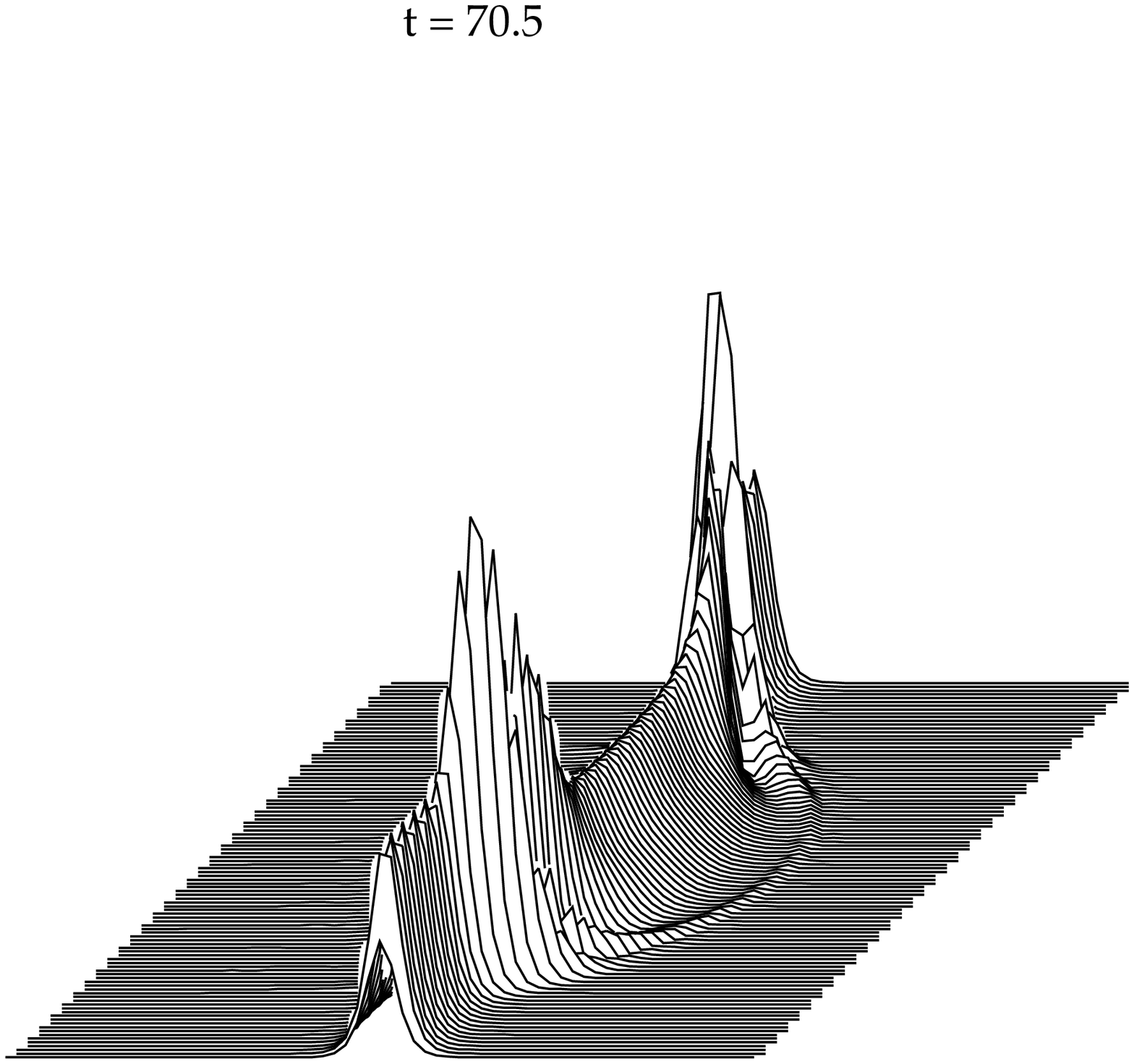}{
Deformation wave : energy density at t = 70.5}

\chapter{Topological waves on the domain wall}

In the previous section, we have derived solutions for which $\psi$ is 
constant. We will now relax this condition and look for waves involving a 
change in $\psi$ propagating on the domain wall \eSinSol.
First of all, looking for solutions of the form 
$f=f(x),\,\, \psi=\psi(y,t)$, we get the following equation of motion:
$$
\eqalign{\psi_{tt}-\psi_{yy}&=0\cr
f_{xx}+{sin2f \over 2} \bigl [(1-f^2_x)(\psi^2_t-\psi^2_y)-\mu^2 \bigr ]&-
\sin^{2}f \bigl [f_{xx}(\psi^2_t-\psi^2_y) \bigr ]=0.\cr}
\eqn \ePsittFxx
$$
Solutions of these equations are given by $\psi(y,t) = \psi(y-t)$ or 
$\psi(y,t) = \psi(y+t)$ and where $f$ is the domain wall solution \eSinSol.
This corresponds to a wave in ``$\psi$'' propagating on the domain wall 
in one direction at the speed of light. 

Let us return to the boundary conditions for the field $\vec \phi$ and
the topological charge.

First of all note that the expression for the topological charge
written in terms of fields $f$ and $\psi$ is given by
$$
N={1 \over 4\pi} \int d(cosf) \wedge d\psi,
\eqn \eTopfpsi
$$
%so that the topological charge carried by the domain wall is given by 
As we have argued before the topological charge carried by our domain wall 
solution \eSinSol\ is zero. This corresponds to $\psi=0$. However, thinking
of the wall as having come from a closed structure (a straight section and 
a ring ``at infinity'') the total charge of this structure can be an integer 
$m$ and the charge carried by the wall-like configurations is given by
$$ 
\Delta \psi \big \vert^{y=+\infty}_{y=-\infty} =2\pi \beta,
\eqn \eDelta 
$$
where $\beta\in Z$ if we impose a further condition that the fields
at $y\rightarrow\pm\infty$ are the same. This last condition is a very
natural generalisation of $\Delta \psi=0$ satisfied by our
domain wall \eSinSol\ and it allows topological waves to propagate along
the wall. 

Consider, for example, $\psi=\pi[\Tanh(y-t)+1]$, taken together with
$f$ given by \eSinSol\ (Fig. 2). This is a solution of \ePsittFxx\ and it 
describes a wave propagating on the domain wall. Its topological charge 
(given by \eTopfpsi\ or \eDelta) is 1. It is important to note that the 
topological charge on the wall does not have to be localised; it can be 
spread out or split into small lumps of fractions of a unit charge.

In Figure 2 we present a snapshot of the solution $\psi = \pi(atan(y-t)+1)$
and where $f$ is given by \eSinSol. The arrows in fig 2.a correspond to a 
projection of the $S^2$
field onto the plane at the equator of the sphere. The orientation of the 
arrows is thus given by $\psi$ and the length of the vectors is given by 
$\mod{sin(f)}$. To distinguish between vectors in the upper or lower 
hemisphere, we add a ``$+$'' or a ``$\times$'' at the origin of the arrows. 

%%%%%%%%%% FIGURE 2 %%%%%%%%%%%%%%%%
\TwoFigsAB{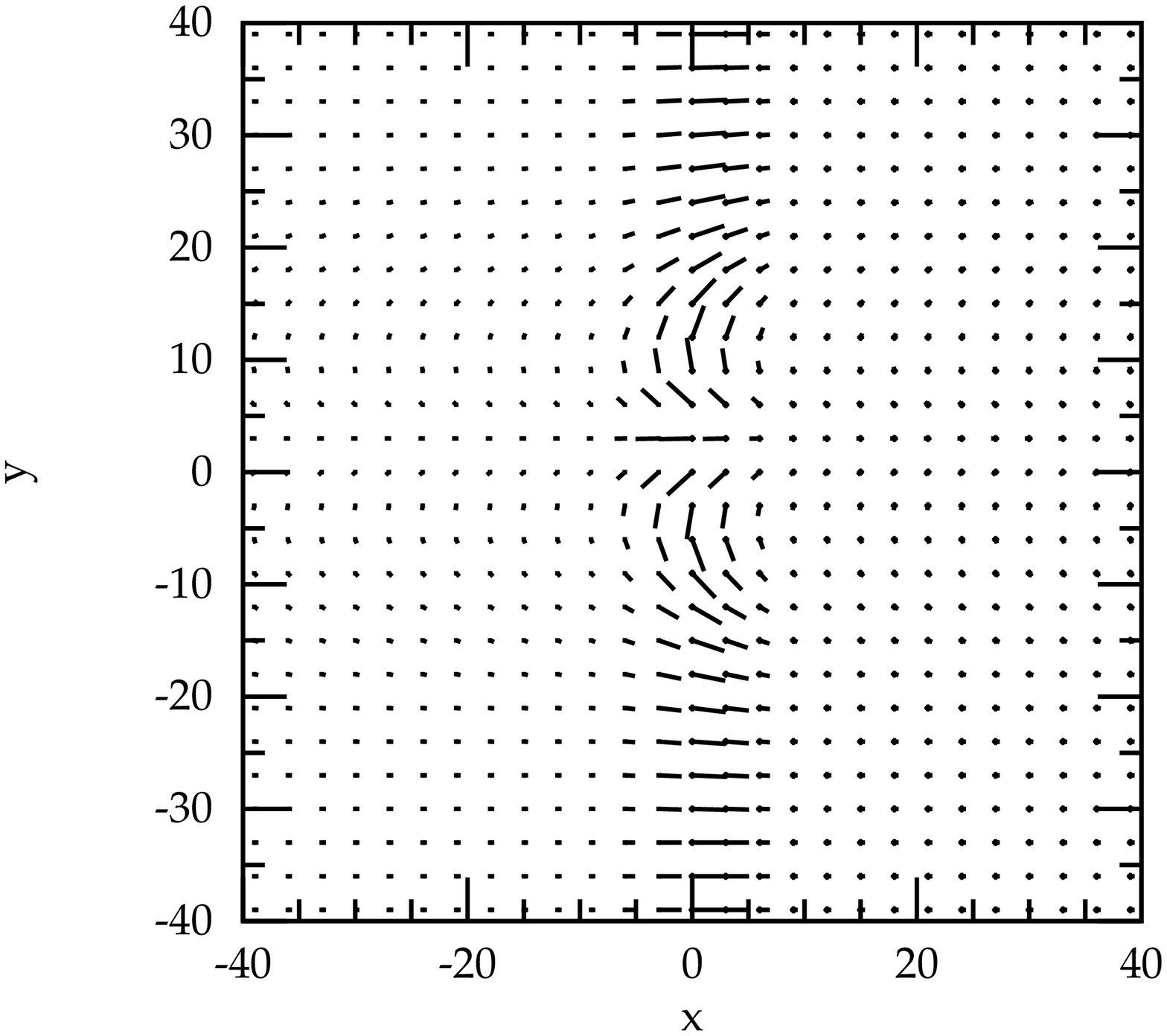}{Topological wave with total charge Q = 1}{
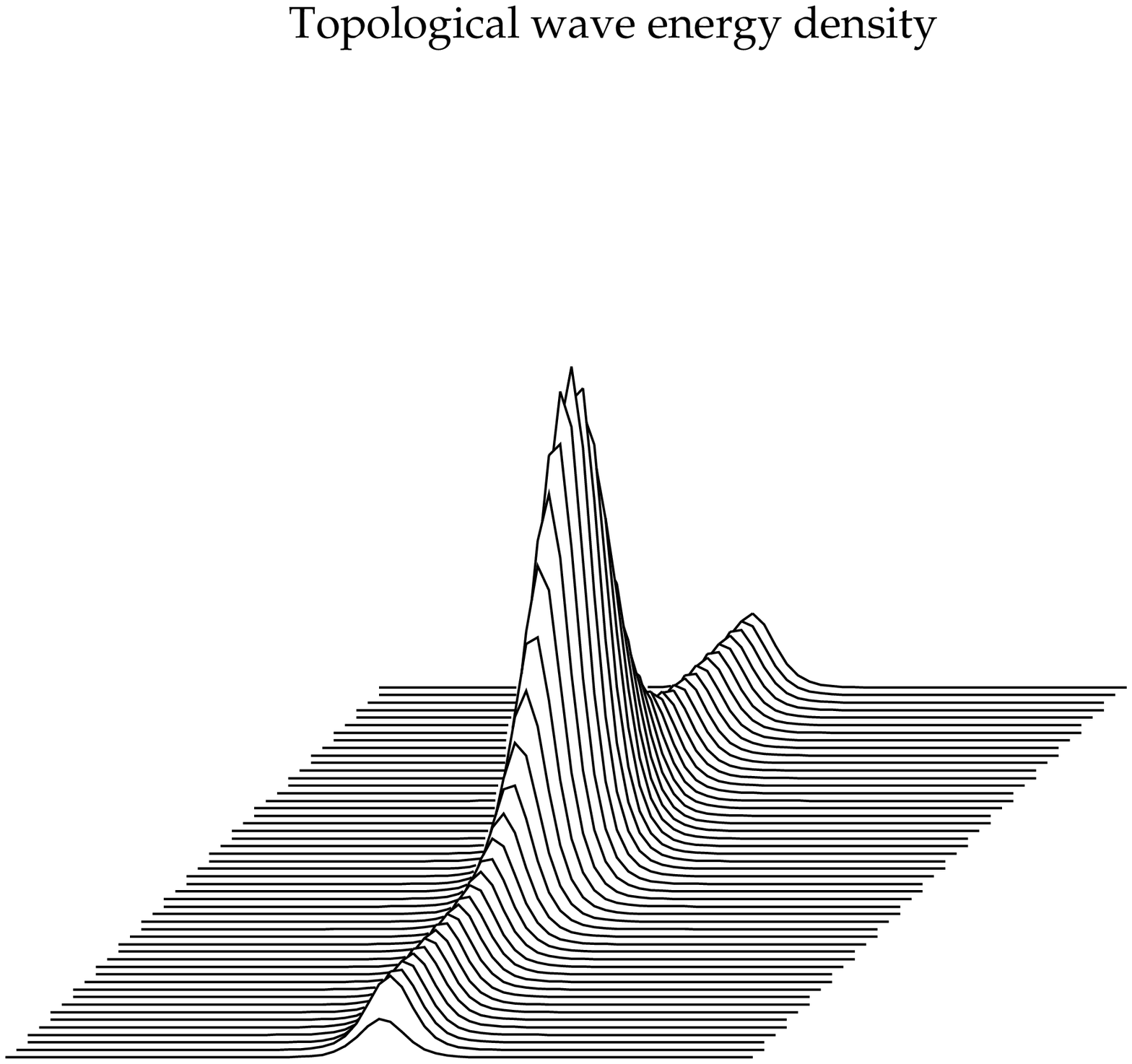}{Energy density for the topological wave}

The collision between two waves
propagating in opposite directions is non-linear and we have solved the
equation of motion numerically to perform the scattering of two such
waves. The collision is inelastic resulting in the emission of a circular wave 
as is shown in Figure 3. In contradistinction to the collision of
two travelling waves this emission is symmetric.

%%%%%%%%%% FIGURE 3 %%%%%%%%%%%%%%
\FourFigsAD{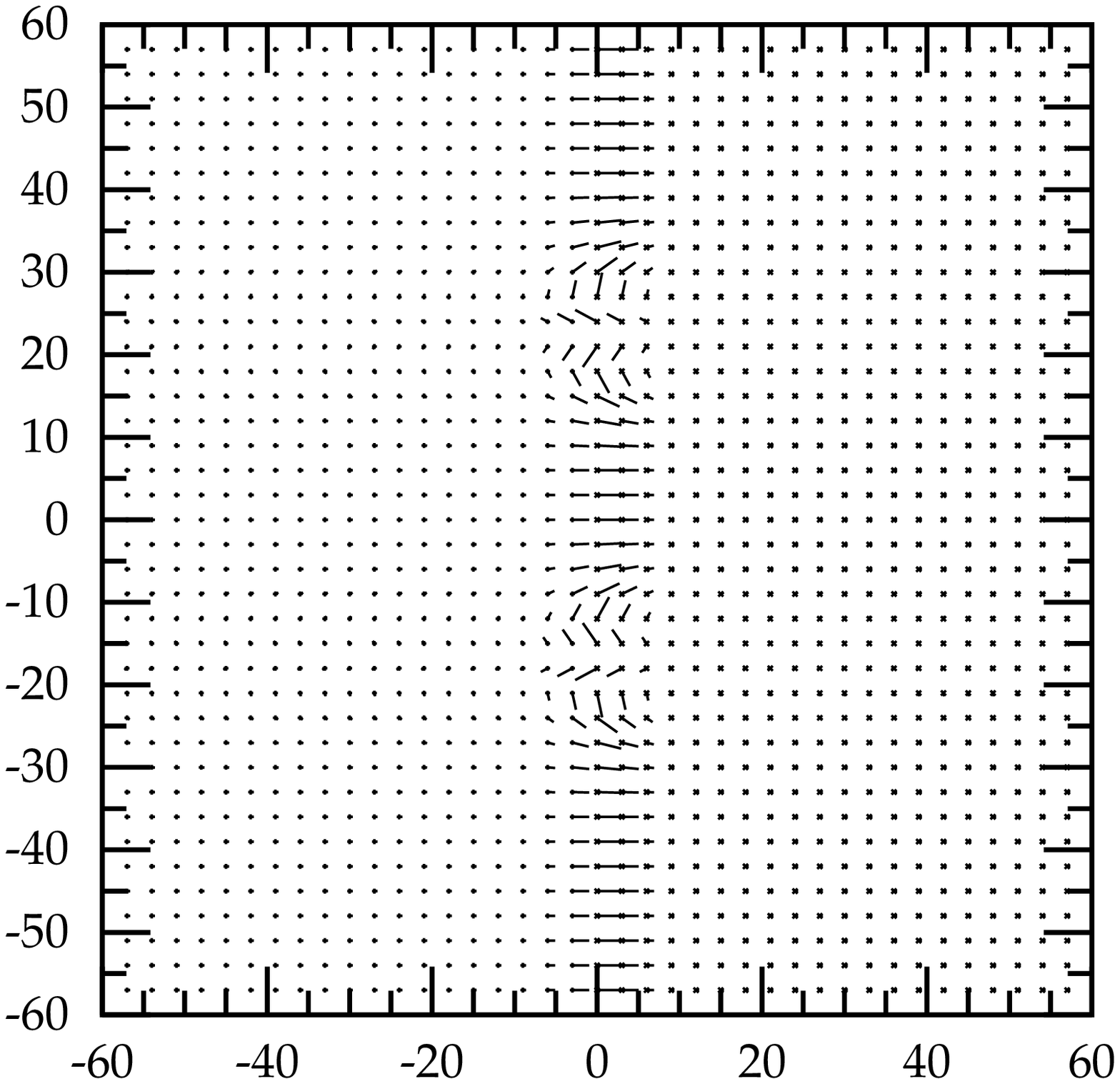}{Field of topological wave scattering at t = 0}{
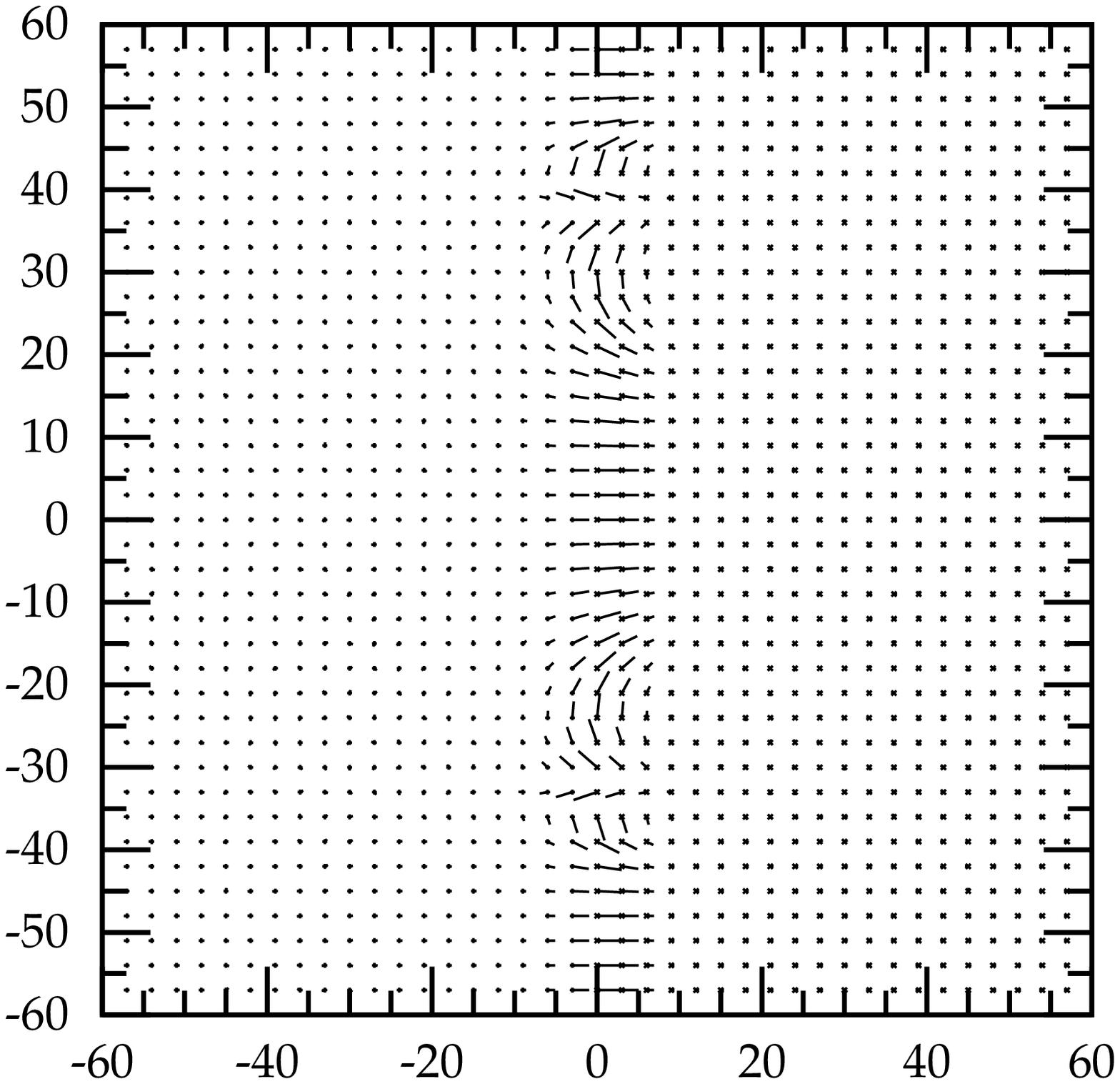}{Field of topological wave scattering at t = 60}{
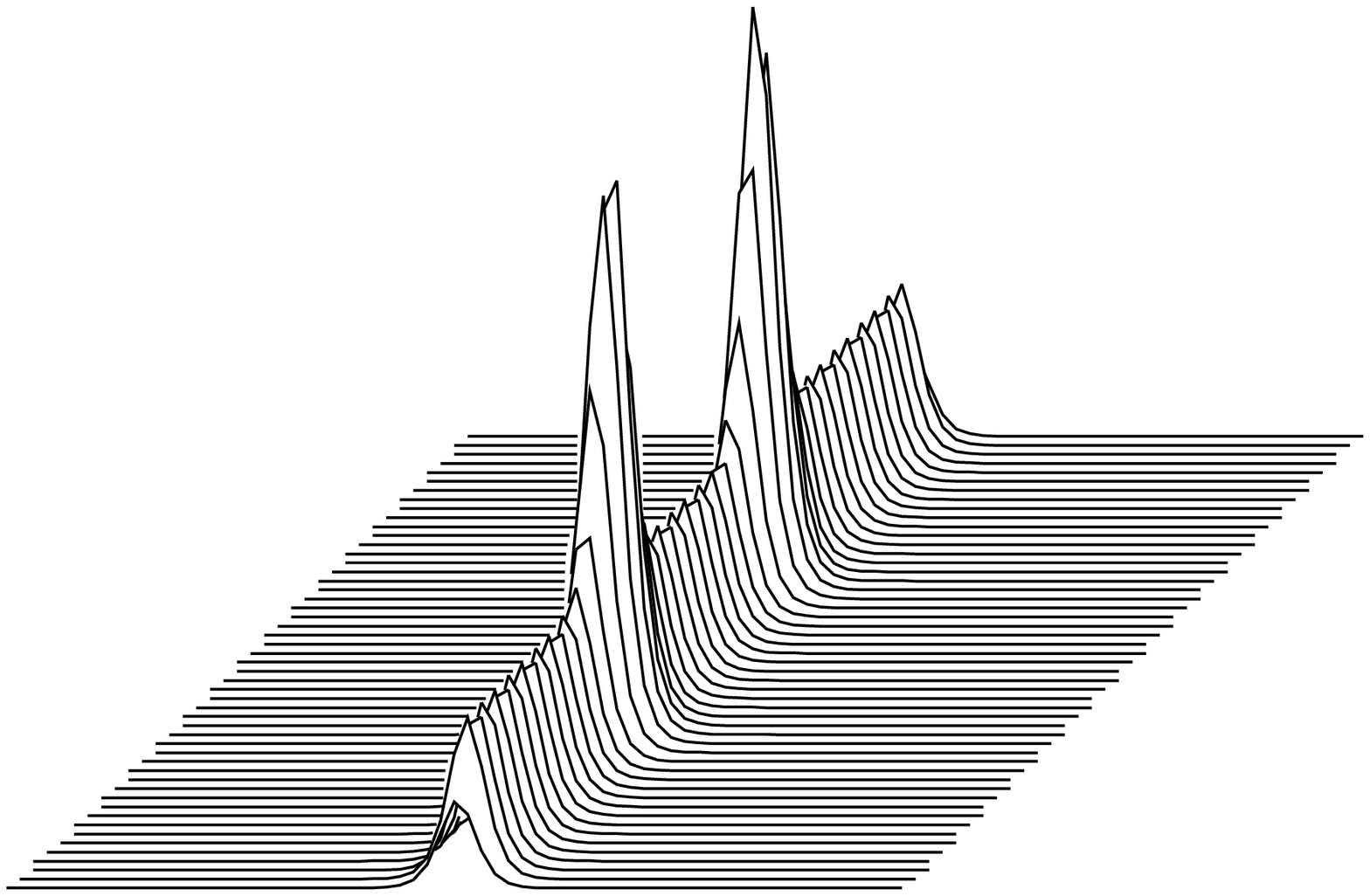}{Energy density of topological wave scattering at t = 0}{
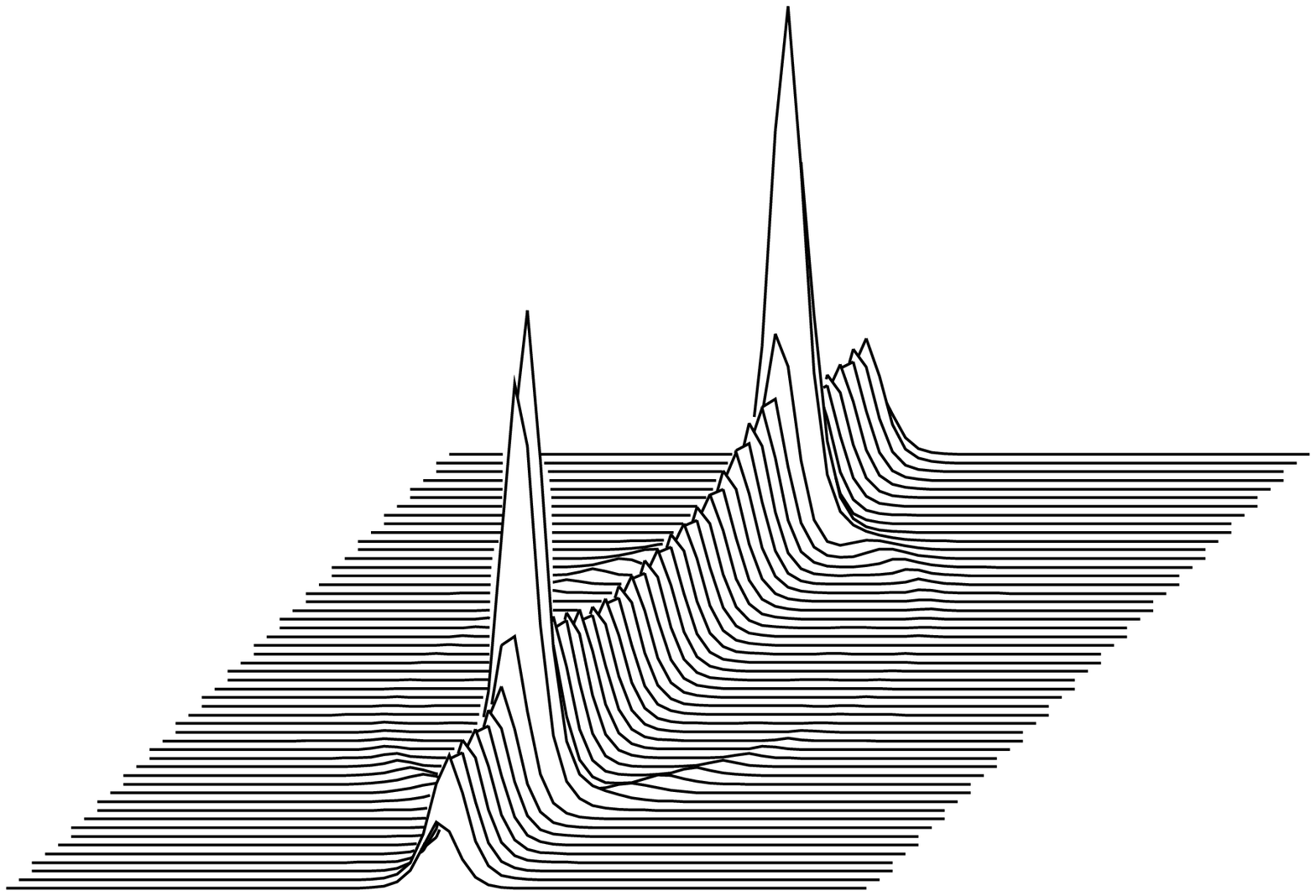}{Energy density of topological wave scattering at t = 60}

Next, we look again for small amplitude waves and try to find solutions of
the form 
$f(x,y,t)=f_s(x) +g(x,y,t), \mod g \ll1$,\quad $\psi=\psi(y,t)$ linearising 
the equation of motion in terms of $g(x,y,t)$. We get
$$
\eqalign{
\psi_{tt}-\psi_{yy}&=0,\cr
g_{tt}-g_{xx}-g_{yy}&+\mu^2(1-{2\over \cosh^2\mu x})g={\sinh\mu x
\over \cosh^2 \mu x} \bigl (1- {2\mu^2 \over \cosh^2 \mu x}\bigr )
(\psi^2_t-\psi^2_y).\cr}
\eqn\eGttnoze
$$
Notice that the right hand side of this equation is orthogonal to any 
zero-mode solution for $g(x,y,t)$ \eGxxNul. So topological waves, when they 
collide, exite only the continuum modes of the soliton. 
This shows that the topological waves and transversal waves \eGzero can
be superimposed. This will play an important role in what follows.

\chapter{Static Skyrmion solution}

An important class of static solutions of the equation 
of motion consists of fields which are invariant under the group of 
simultaneous spatial rotations by an angle $\varphi \in [0,2 \pi]$ and 
iso-rotations by $-n \varphi $, arround the $\phi_3$  axis, where $n$ is a 
non-zero integer. Such fields are of the form 
   
$$ 
\vec \phi(\vec x)= \left(
\matrix{ &\sin f(r) \cos (n\theta)\cr
 &\sin f(r) \sin(n\theta)\cr
 &\cos f (r)\cr}
\right),
\eqn\eHedgehog
$$                         
where $(r,\theta)$ are polar coordinates in the $(x,y)$-plane.
Such fields are analogues of the hedgehog field of the Skyrme model and were 
studied in [\RNo-\RNT] for the baby-Skyrmion model.

The function $f(r)$, the
analogue of the profile function of the Skyrme model, has to satisfy
$$ 
f(0)=m\pi ,\quad  m \in Z
\eqn\efzero
$$                                  
for the field \eHedgehog\ to be regular at the origin. As we are looking for 
field configurations with a finite energy we set
 $$ 
lim_{r \to \infty}f(r)=l\pi,~~~ l \in Z. 
\eqn\efinfinity
$$
Clearly, Skyrmion solutions with $l=2k,~~ k \in Z$ at $r \to \infty$ approach 
the $\phi_3=\phi_+$ vacuum state and those with $l=2k+1,~~ k \in Z $ go to 
the $\phi_-$ vacuum state.
The boundary conditions \efzero\ and \efinfinity\  are responsible for making 
this class of configurations topologically nontrivial. 
Moreover, the degree of the fields \eHedgehog\ (their winding number) is:   
$$        
deg[\vec \phi]=[\cos f(\infty)-\cos f(0)]{n \over 2}=\bigl [(-1)^l-(-1)^m 
\bigr ]{n \over 2}. \eqn\eDegree
$$

The $Z_2$ symmetry of the model transforms the two types of Skyrmions into one 
another. We can thus, without any loss of generality, set $l$ to $0$.
For the case $n=1,~~ m=1,~~ l=0,$ we have $deg[\vec \phi]=1$. 
The Euler-Lagrange equation for $f (r)$ is given by
$$ 
(r+{n^{2}\sin^{2}f\over r})f^{\prime\prime}+
(1-{n^{2}\sin\sp2 f \over r^{2}} +{ n^{2}f^\prime \sin f\cos f \over r})
f^\prime-{n^2\sin f\cos f \over r} -r{{\mu}^{2} \over 2 }\sin 2f=0.
\eqn\efeqn
$$
   
The profile function $f(r)$ for the case $n=1,~~ m=1$ and $l=0$ is shown in 
Figure 4 (Skyrmion). The case
$m=1, l=0$ and $n=-1 $ corresponds to $deg[\phi]=-1$, has the same profile 
function $f$ but it describes an anti-skyrmion. We have found numerically 
that the total energy of the Skyrmion solution is $ E_{Sk}=1.446$
in our units. This value is smaller than 
the energy of the baby-Skyrmion ($E=1.56$) found in [\RNo].

%%%%%%%%%%%% FIGURE 4 %%%%%%%%%%%%%
\TwoFigsAB{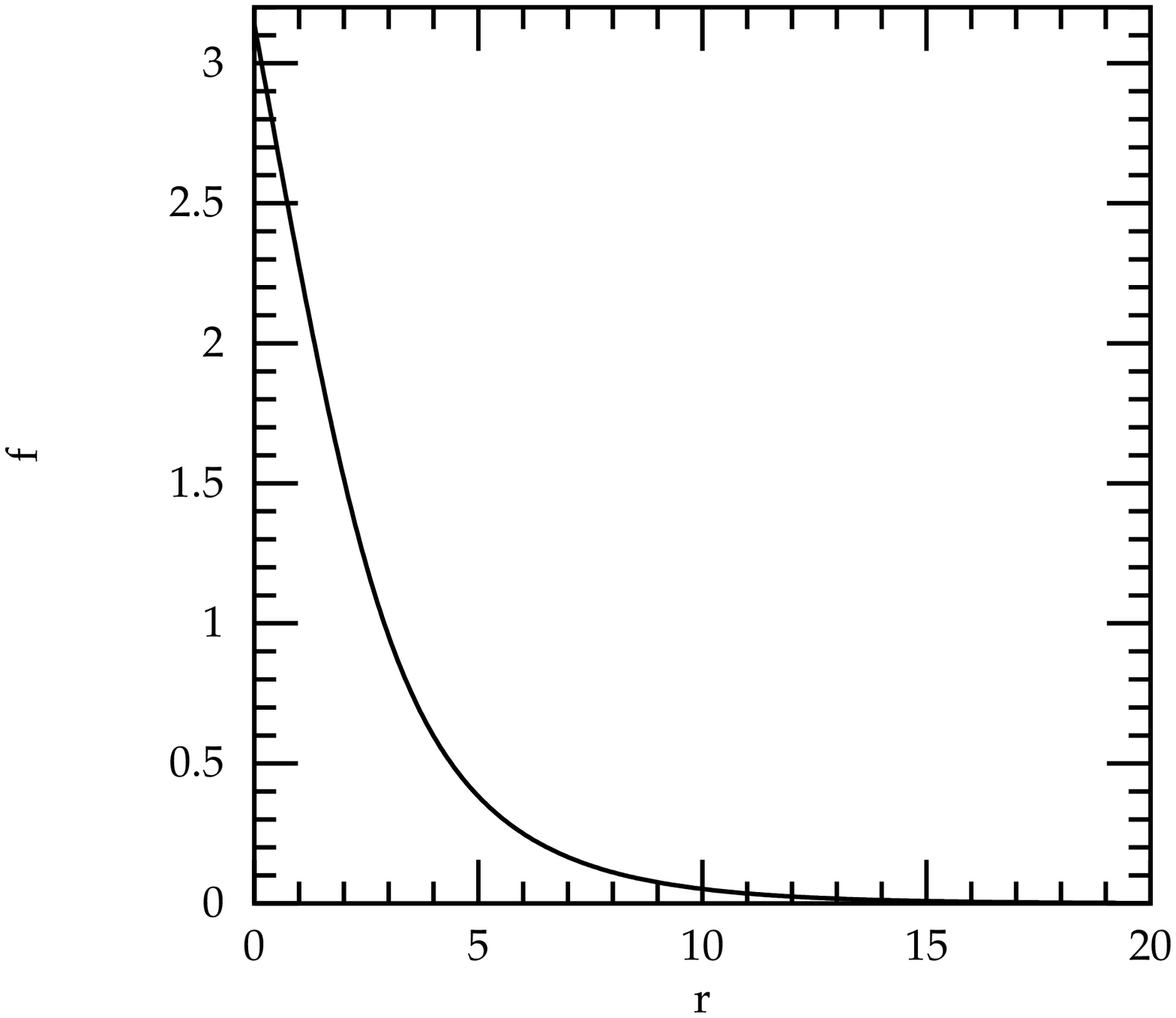}{Profile function $f$ for the Skyrmion}{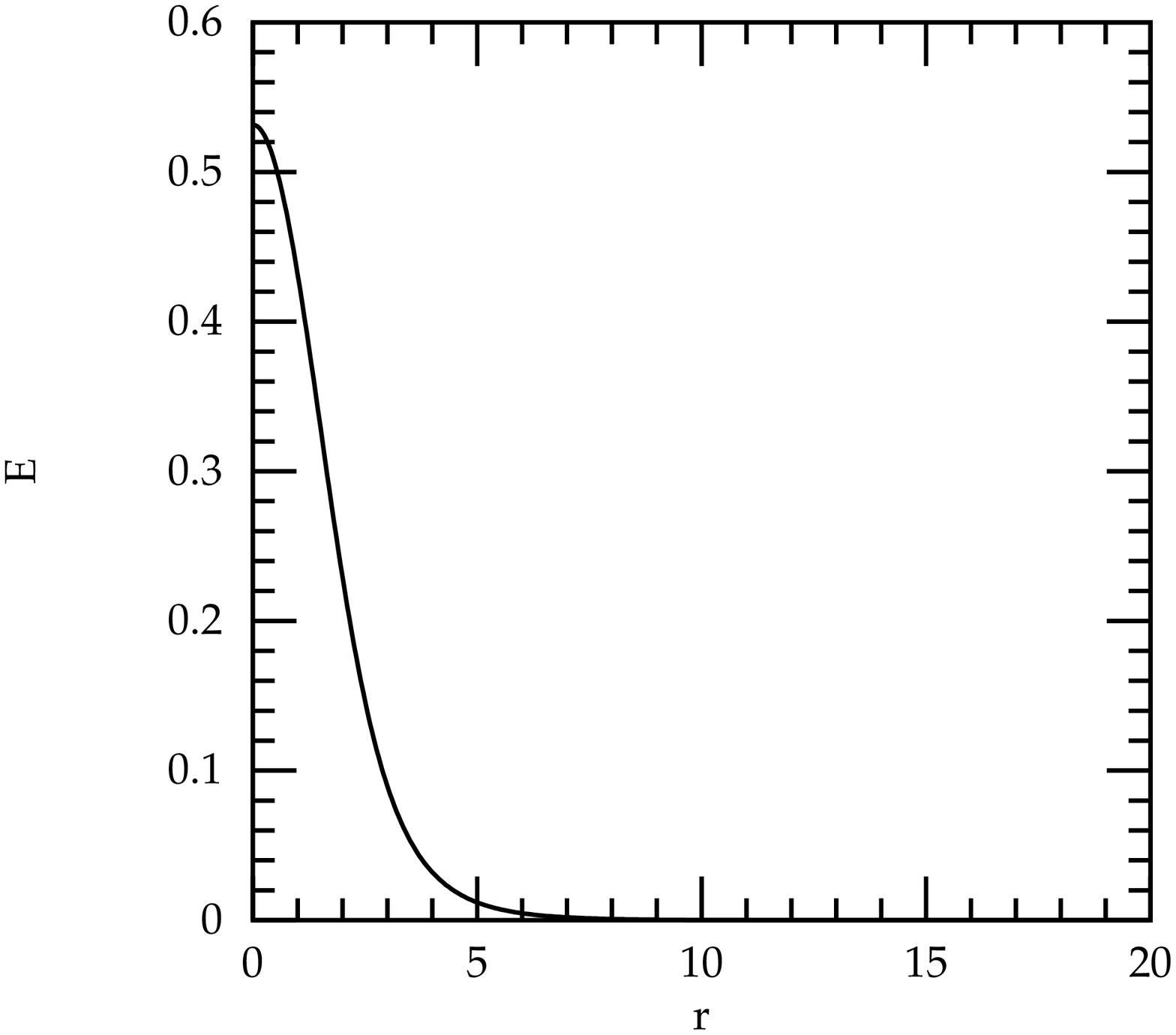}{
Energy profile for the Skyrmion}

\chapter{Skyrmion absorption by domain walls.}

So far we have seen that the model described by \eLagPhi\ has soliton-like
solutions corresponding to domain walls, as well as topological
solitons (Skyrmions). We will now study the interaction between these two 
types of solutions. The force between a domain wall and a Skyrmion 
depends both on the distance between them and their relative orientation.
To superimpose two solutions, we use a stereographic projection 
of the sphere onto the complex plane and use the field $w$ defined as
$$
  w = {\phi_1 + i\phi_2 \over 1 + \phi_3}.
\eqn\ewMap
$$
In this formulation, the vacuum configuration $\phi_{+}$ is given by $w = 0$
and we can in first approximation, superimpose a Skyrmion and
the domain wall (on the $\phi_{+}$ side) by adding the 2 fields configurations.
As we must also choose a relative orientation between the two structures we
can write:
$$
	w = w_{wall} + e^{i \alpha \pi} w_{Sk},
\eqn \ewSupp
$$
where $w_{wall}$ and $w_{Sk}$ are centered arround two points separated by
a distance $R$.
In Figure 5 we present a plot of the total energy of the Skyrmion-wall 
configuration
as a function of the separation distance $R$ for different relative 
orientations
$\alpha$. The energy increase at short distances is due to the fact that 
at these distances the
superposition of the two solutions is not a good approximation to the true 
configuration.
When $\alpha = 0$ the Skyrmion and the wall are in the
repulsive channel and they repel. When $\alpha = 1$ the wall attracts the
Skyrmion and, we have checked numerically that for other values of $\alpha$,
the Skyrmion and the wall rotate themselves into the attractive channel
and move towards each other. 

%%%%%%%%%%%% FIGURE 5 %%%%%%%%%%%
\OneFig{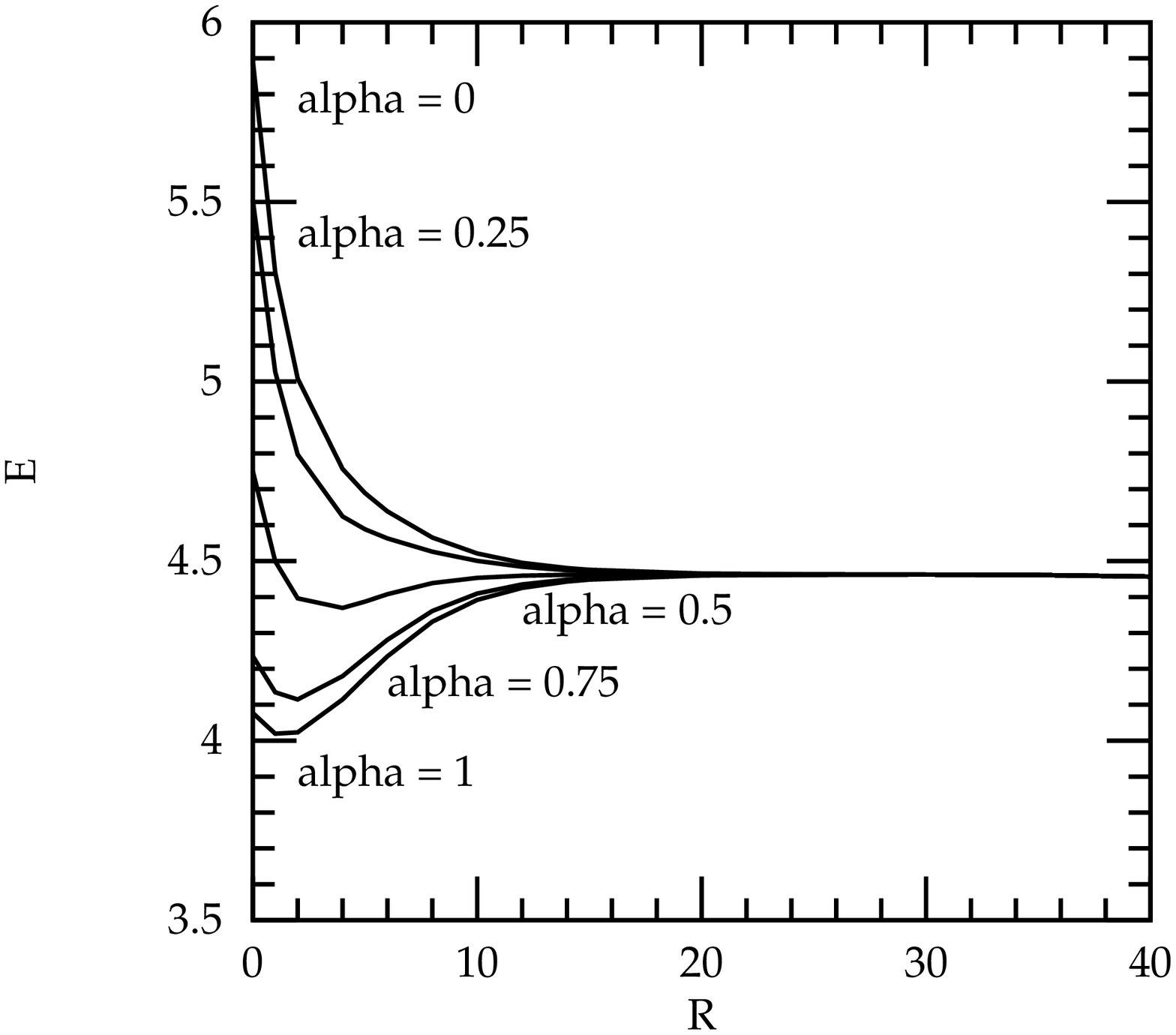}{Skyrmion-wall potential for $\alpha = 0, 0.25, 0.5, 
0.75$ and $1$.}

After the Skyrmion has collided with the wall it splits into two parts
which propagate along the wall at the speed of light (Figure 6). 
Each part carries one half of the unit of topological charge. 
These two waves are very similar to the superposition of 
topological and deformation waves given by the solutions of \eGttnoze
The collision is inelastic in the sense that a circular wave is produced
in the $(x,y)$ plane implying that the reverse process, of creating a Skyrmion
from the wall, is not easy.

%%%%%%%%%%%%% FIGURE 6 %%%%%%%%%%%%
\FourFigsAD{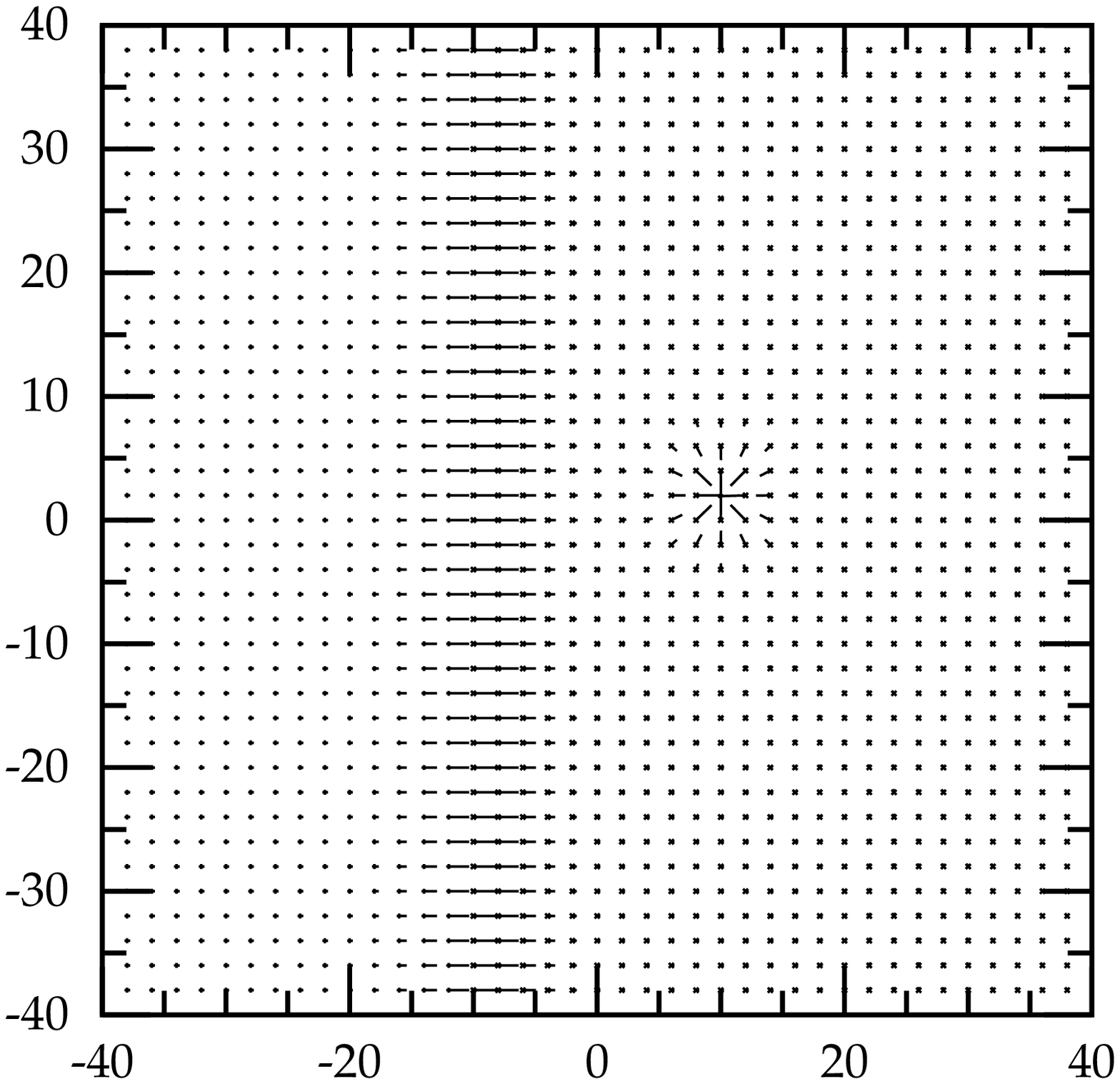}{Skyrmion-wall field at t = 0}{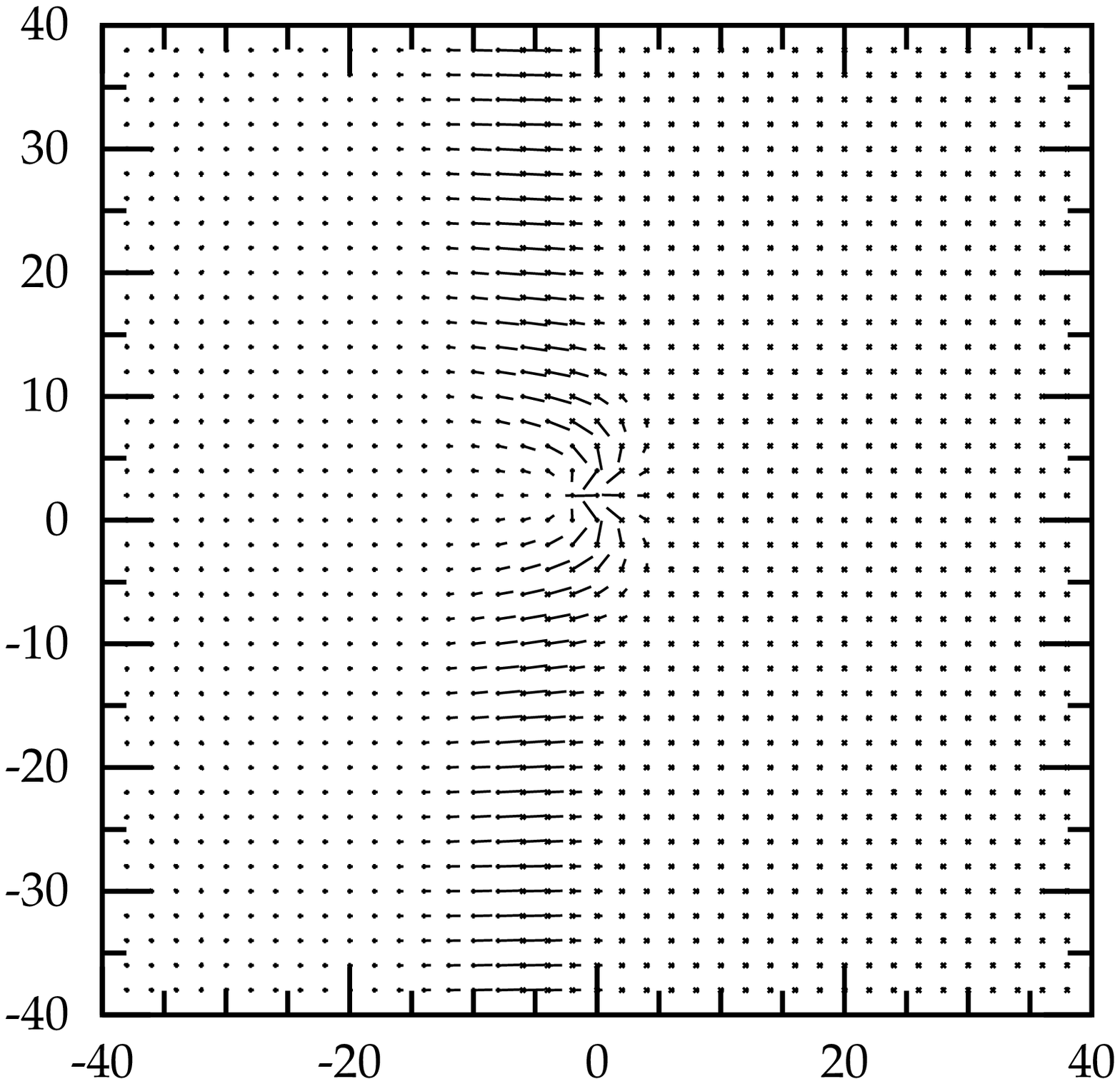}{
Skyrmion-wall field at t = 123}{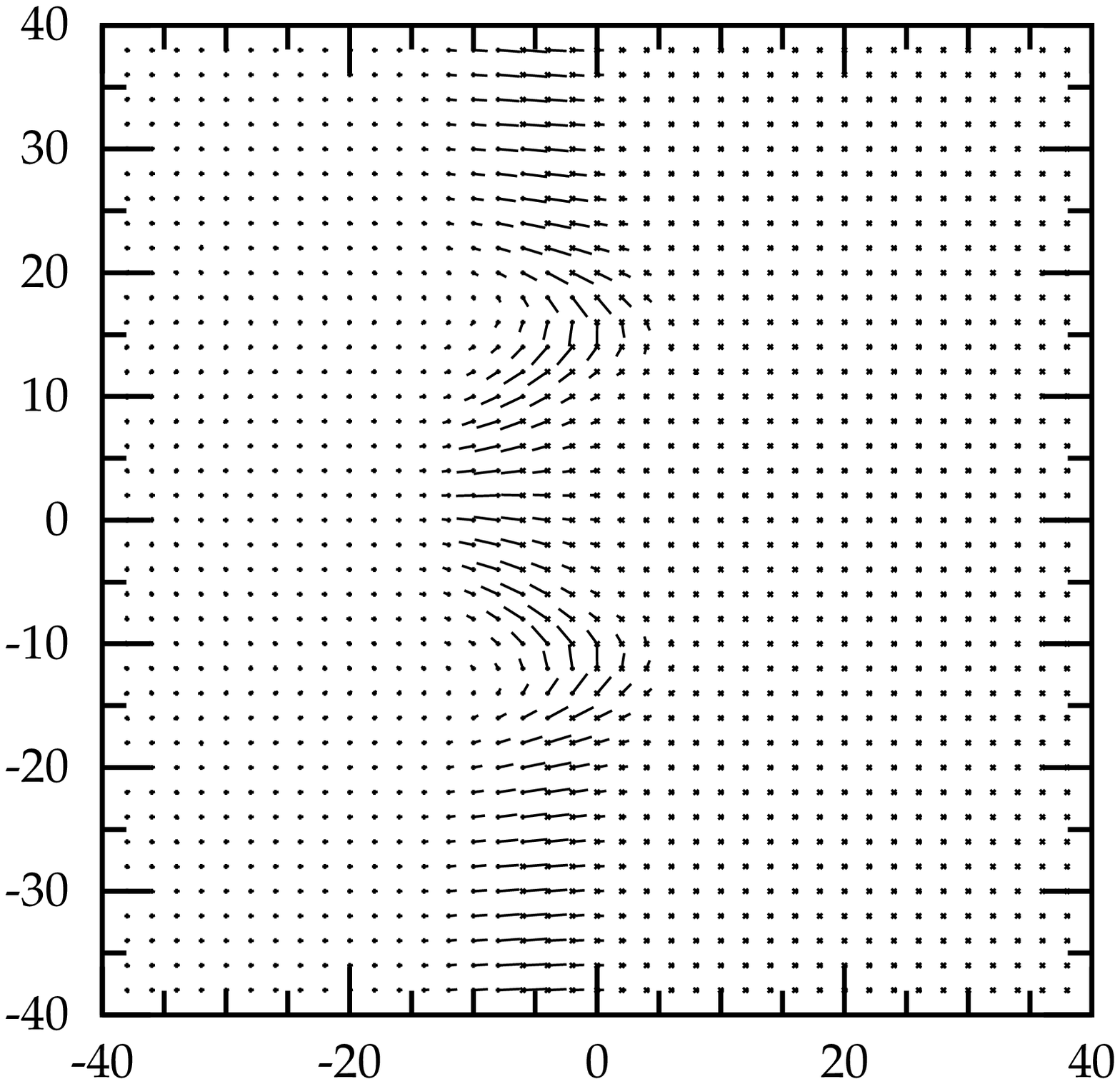}{Skyrmion-wall field at t = 135}{
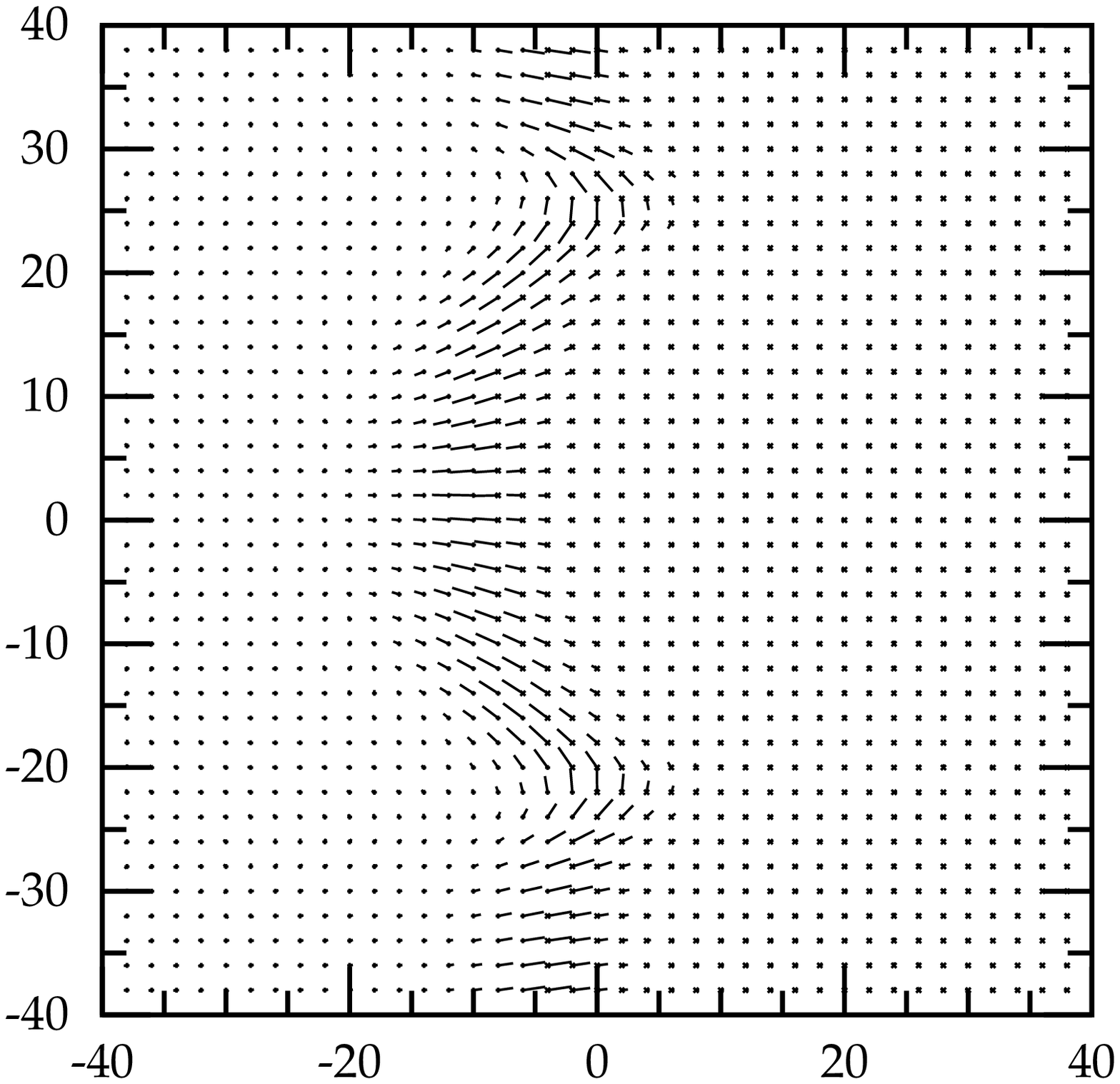}{Skyrmion-wall field at t = 145}

\chapter{Creation of a Skyrmion by a domain wall}

A domain wall is the lowest energy configuration of the  $\vec \phi$-field 
corresponding to the appropriate boundary conditions.
This is why only some excitations of the wall produce particles or waves in 
the $(x,y)$-plane. We have already mentioned that the interaction of 
topological wave packets is a possible source for such a process.

As our first attempt to create a Skyrmion from the wall we put two 
Skyrmions at a finite distance from each other and made them be absorbed
by the same wall at the same time. Each Skyrmion produced two topological 
waves on the wall; two of these waves collided with each other, 
creating what looked like a Skyrmion, but there was not enough energy to
emit it from the wall.  

Looking at the topological wave produced by a Skyrmion when absorbed by a 
wall, it is clear that we have to combine a topological lump with a 
deformation wave as described in the third section. 
Moreover, topological waves are necessary to give a topological 
charge to the Skyrmion. 
On the other hand, deformation waves do not carry any topological 
charge, but these travelling zero-mode play an exceptional role 
among all  excitations of the wall, namely, they preserve the information 
about the initial deformation of the wall. This is why they can be a source 
of the energy transfer from the $y$-direction to the $x$-direction. This is 
clearly seen in Figure 1, which shows that such a collision process is 
strongly anisotropic. The emission of  waves in this case is mainly directed 
to the right from the $y$-axis. We see in this figure that at a certain time, 
a jet-like configuration, with an energy peak being emitted along the positive 
$x$-axis, is formed. The process is very asymmetric; the overall momentum is 
conserved by a (small) recoil of the domain wall. Of course, for an infinitely 
long wall this recoil is negligible.

To create a Skyrmion, we have taken the following initial condition
corresponding to the superposition of a deformation and a topological
wave  
$$
\eqalign{
   w = \exp\Bigl(-\mu \bigl(x- {B\over 2}
       [\Tanh(\mu(y + t - A)) - \Tanh(\mu(y - t + A))]\bigr)\bigr) \cr
        \exp\Bigl({i \over 2} \pi\bigl(\Tanh(\mu(y + t - A - D))+ 
                               \Tanh(\mu(y - t + A + D))+2\bigr)
                \Bigr)\cr
}
\eqn\eEmitSk
$$
where $A$ sets the initial distance between the waves packets, 
$B$ corresponds to the amplitude of deformation (kink) and $D$ introduces 
a delay between the deformation wave and the topological wave on each packet.
(The initial condition is actualy given by $w$ and ${d w\over d t}$ both 
evaluated at $t=0$).
When $D > 0$ the topological wave is retarded with respect to the deformation 
wave. This is the type of configuration observed in Fig 6d after the wall 
has absorbed a Skyrmion. 
For the numerical simulations, we chose $A = 30, B = 20, D = 10$. 
As is shown in Figure 7, this initial condition results in the emission of
a Skyrmion at the right hand side of the wall. After being emited the 
Skyrmion moves away from the wall at about half the speed of light and 
as we can see from the figure 
the distance between the wall and the emited Skyrmion is larger than 20.
This, together with the potential shown on Figure 5, clearly indicates that
the Skyrmion is detached from the wall.

%%%%%%%%%%%% FIGURE 7 %%%%%%%%%%%%
\FourFigsAD{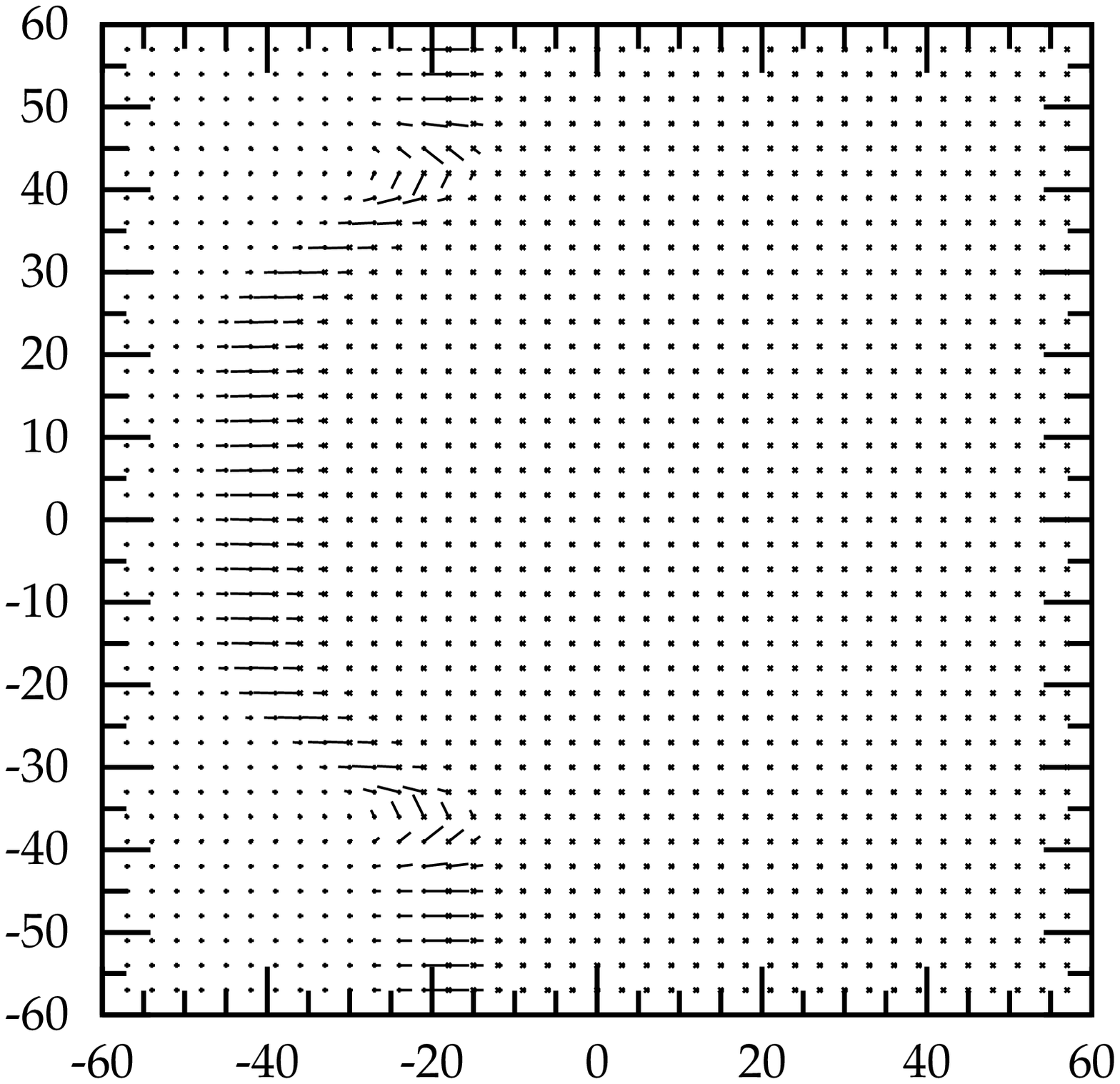}{Skyrmion emission.\break Field at t = 0}{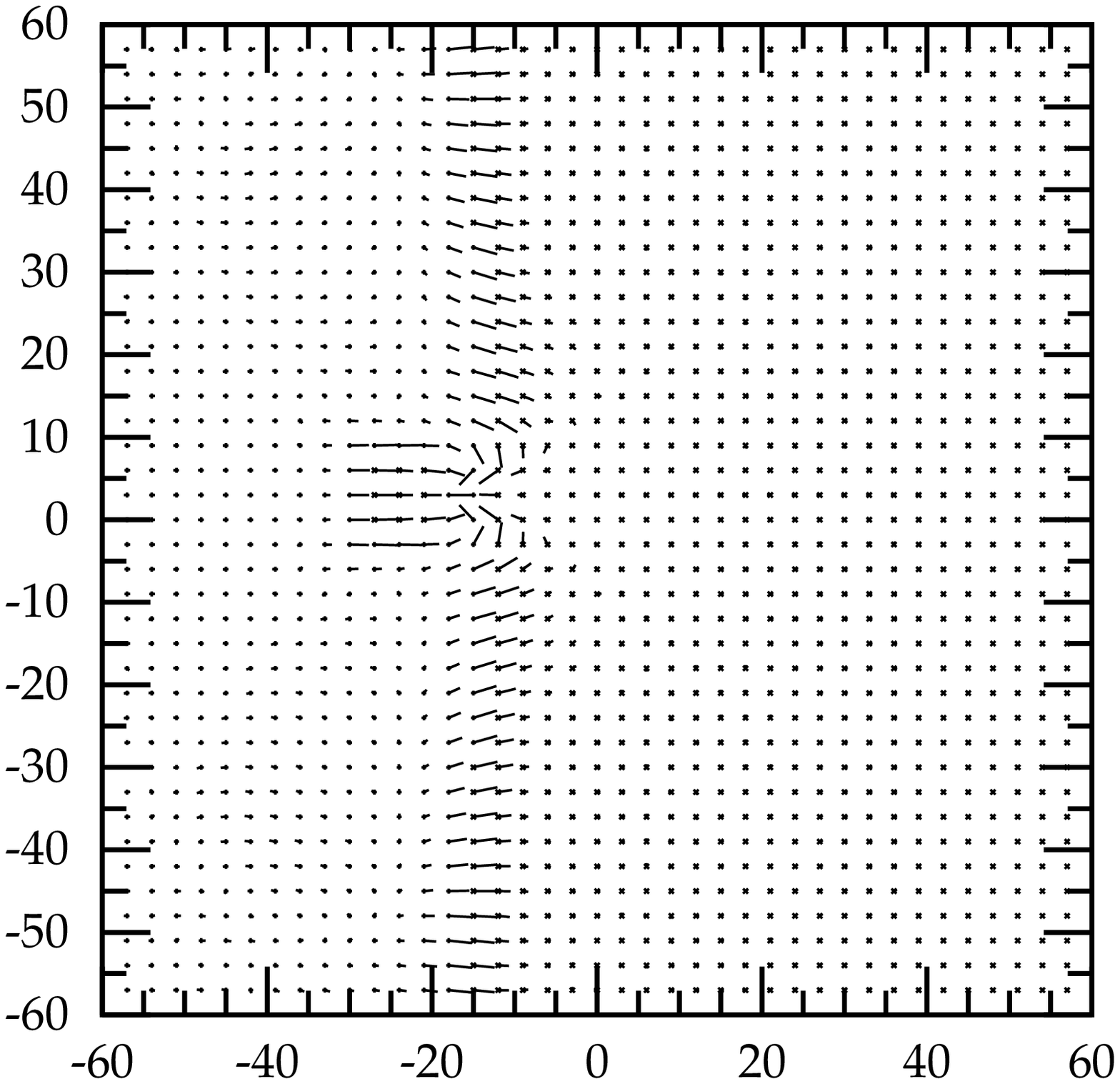}{
Skyrmion emission.\break Field at t = 36}{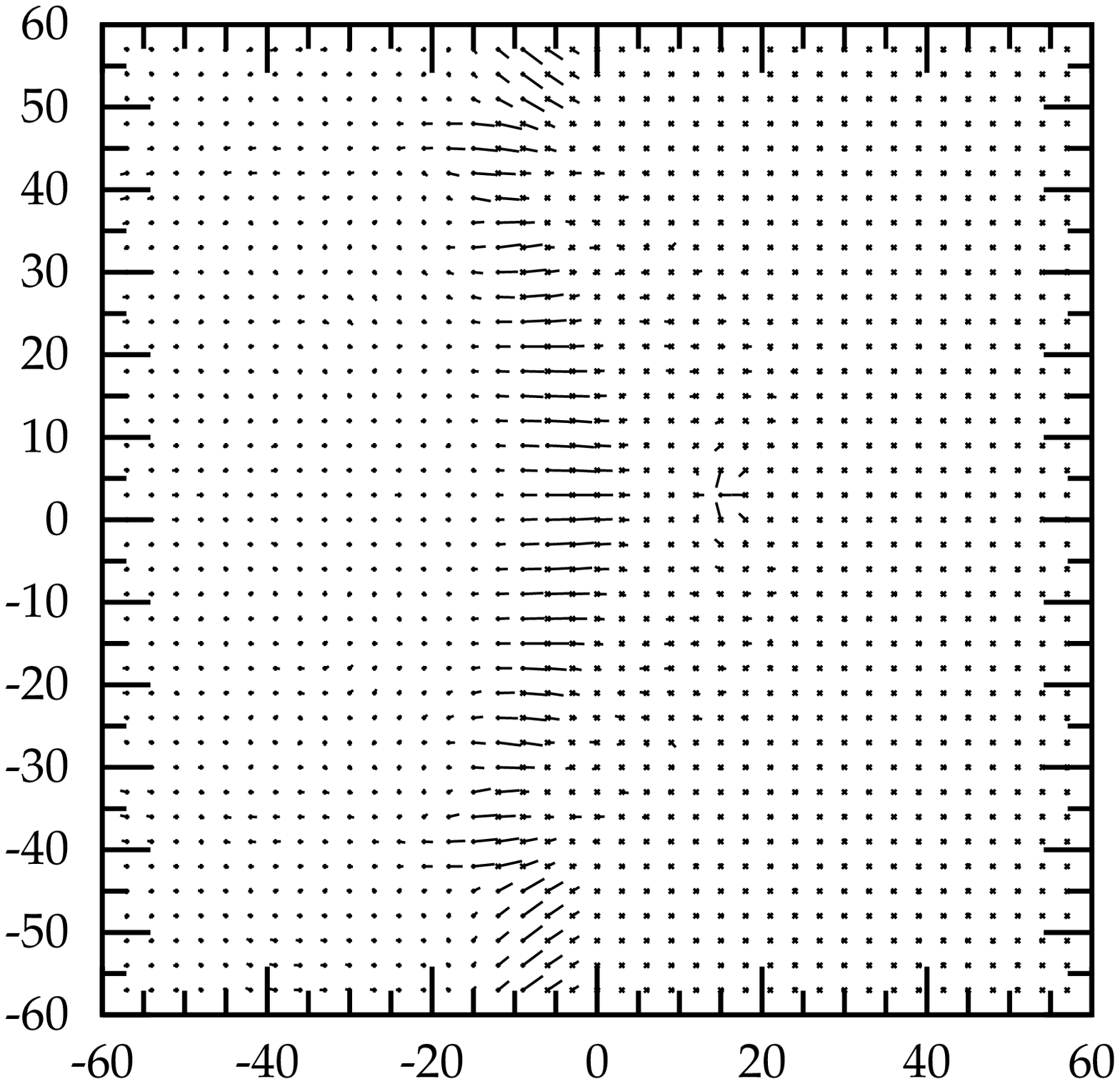}{Skyrmion emission.
\break Field at t = 79.5}{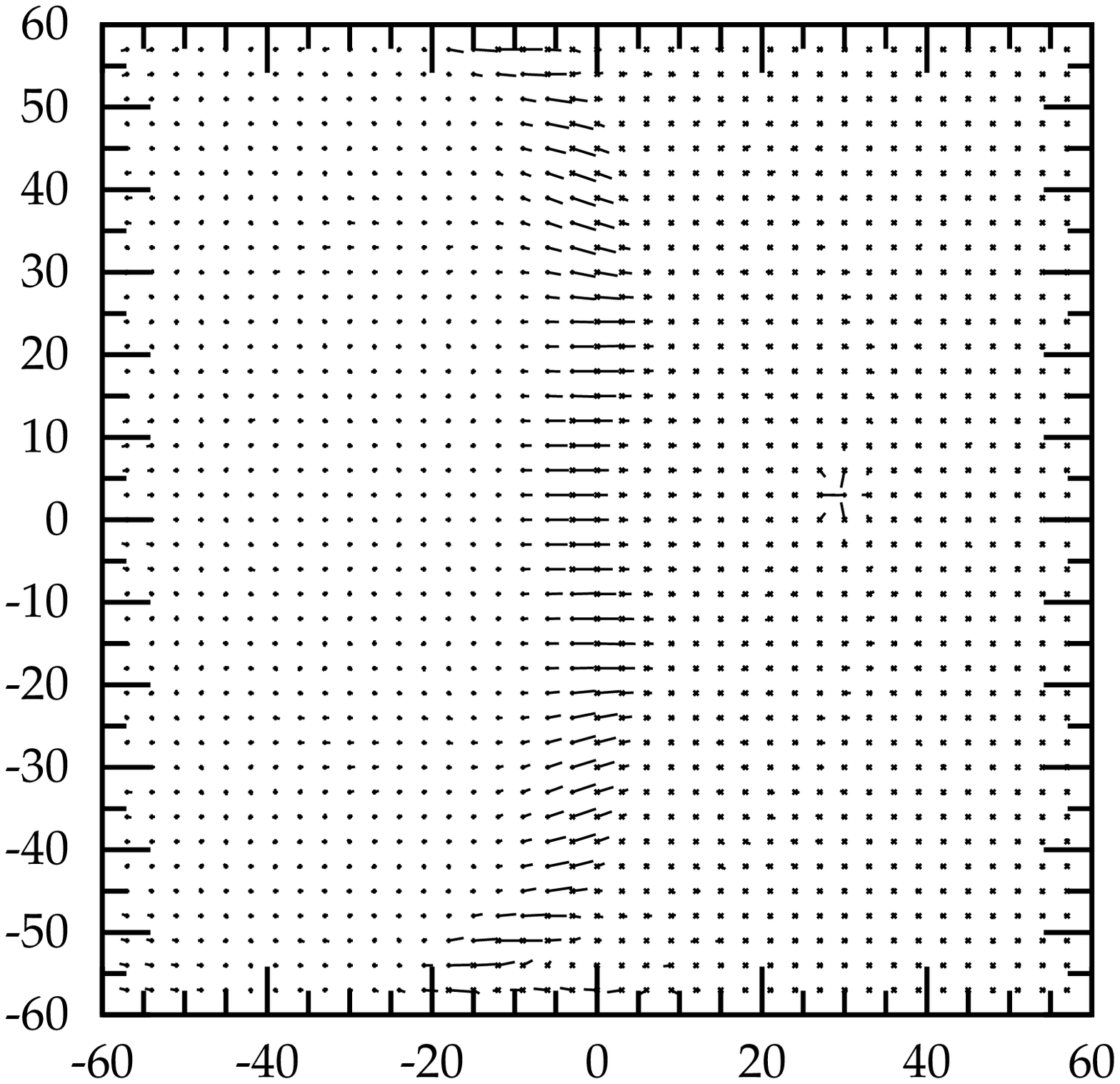}{Skyrmion emission.\break Field at 
t = 102}

We have also used other values of $D$. When $D$ is negative, the Skyrmion
is emitted on the other side of the wall, but both the wall and the Skyrmion
move to the right after the emission.
This suggests that the
wall might reabsorb the Skyrmion after a while (something we are not able
to show numerically). When $D = 0$ the wall does not emit any Skyrmion.

The plots in Figure 7 clearly confirm the emission of an isolated Skyrmion 
by the wall. Moreover, the process is inelastic, i.e. the production of 
a Skyrmion is accompanied by the emission of mesonic waves.

It is also worth mentioning that in the collision process described above the 
Skyrmion is created not in its ground state but rather in an exited state. 
Consequently, the height of the energy density peak of the Skyrmion 
oscillates around its  position of  equilibrium (after emission).

\chapter{Conclusions}
We have shown that the model described by the Lagrangian density  \eLagPhi\
has various types of classical solutions including topological solutions
(Skyrmions) and domain walls. The domain wall acts as a carrier of
waves propagating at the speed of light. 
Some of these waves have a non-zero topological charge.

Skyrmions and domain walls attract each other leading to the absorption
of Skyr-mions by the walls and the creation of topological waves.
By choosing appropriate initial conditions it is possible to 
create a Skyrmion from a domain wall. This production takes place for a 
relatively large range of parameters of these initial conditions.
 
We hope to be able to apply the observed phenomena to problems of modern 
high energy physics. The most interesting applications would involve problems 
of baryogenesis, (see, e.g. paper [\RNoo,\RNot] in this connection).

\ack
One of the authors (AK) thanks the Department for Mathematical Sciences of 
University of Durham for the hospitality during his visit. This visit was 
supported partly by a UK Royal Society grant,
by INTAS grant 93-633 -EXT and partly by grant 
 RFFR-95-02-04681.

The authors want to thank J. Dziarmaga for bringing  to their attention the 
papers [\RNoT,\RNof] and him and R.A. Gregory and R.S. Ward for useful comments..

\refout 
 
\end